\documentclass[aps, 10pt, pra, twocolumn]{revtex4-1}

\usepackage{amsmath,xcolor}
\usepackage{amsfonts}
\usepackage{amssymb}
\usepackage{graphicx}
\usepackage{srcltx}
\usepackage{mathcomp}   
\usepackage{ulem}       
\usepackage[mathscr]{euscript}
\usepackage{mathrsfs}
\usepackage{tabularx,ragged2e,booktabs}

\newcommand{\beq}{\begin{equation}}
\newcommand{\eeq}{\end{equation}}
\newcommand{\bea}{\begin{eqnarray}}
\newcommand{\eea}{\end{eqnarray}}
\newcommand{\ket} [1] {|#1\rangle}  
\newcommand{\bra} [1] {\langle#1|}  
\newcommand{\braket}[2]{\langle #1 | #2 \rangle}    

\begin{document}

\title{Practical security bounds against the Trojan-horse attack in quantum key distribution}

\author{M. Lucamarini$^{1,2}$}
\author{I. Choi$^{1}$}
\author{M. B. Ward{$^{1}$}}
\author{J. F. Dynes$^{1,2}$}
\author{Z. L. Yuan$^{1,2}$}
\author{A. J. Shields$^{1,2}$}

\affiliation{\bigskip $^1$Toshiba Research Europe Ltd, 208
Cambridge Science Park, Cambridge CB4 0GZ, United Kingdom}
\affiliation{$^2$Corporate Research $\&$ Development Center,
Toshiba Corporation, 1 Komukai-Toshiba-Cho, Saiwai-ku, Kawasaki
212-8582, Japan}

\begin{abstract}
\noindent In the quantum version of a Trojan-horse attack, photons are injected into the optical modules of a quantum key distribution system in an attempt to read information direct from the encoding devices. To stop the Trojan photons, the use of passive optical components has been suggested. However, to date, there is no quantitative bound that specifies such components in relation to the security of the system. Here, we turn the Trojan-horse attack into an information leakage problem. This allows us quantify the system security and relate it to the specification of the optical elements. The analysis is supported by the experimental characterization,  within the operation regime, of reflectivity and transmission of the optical components most relevant to security.
\end{abstract}

\maketitle

\section{Introduction}
\label{sec:1}

\noindent Since ancient times, the Trojan-horse has been known as a stratagem for penetrating a securely protected space. It is therefore essential to consider Trojan-horse attacks in determining the boundaries of any supposedly secure space. This explains their ubiquitous presence in different fields where privacy is required, ranging from cryptography to computing and finance. In particular for a cryptographic application like quantum key distribution (QKD)~\cite{BB84,GRT+02,SBC+09,LS09}, as well as for its most recent developments showing full or partial independency from the specific devices used~\cite{E91,MY98,BHK05,ABG+07,BP12,LCQ12}, the existence of a protected area is a fundamental assumption.

\begin{figure}[tbp]
  \includegraphics[width=8.5cm]{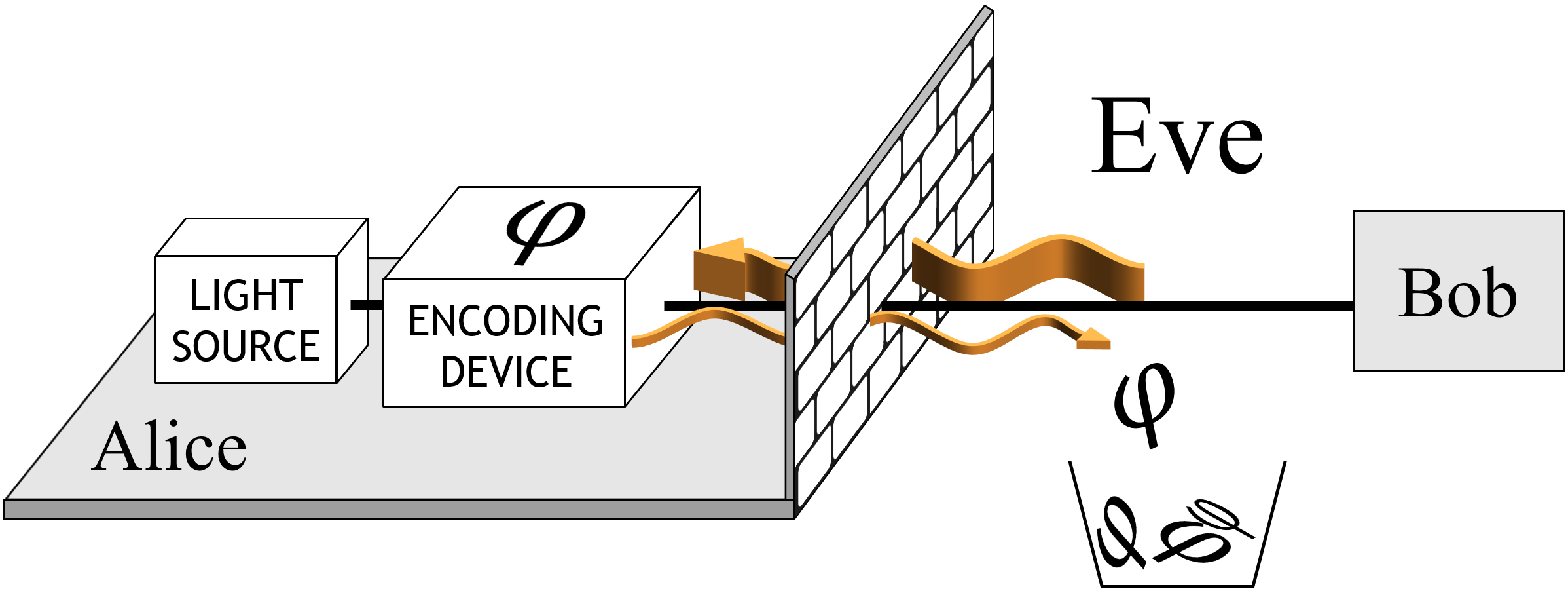}\\
  \caption{Representation of the Trojan-horse attack against an optical QKD setup. Eve sends a large amount of Trojan photons (thick arrow) against Alice's defensive structure. Some of the photons reach the encoding device, are encoded with the private information $\varphi$ and reflected back to Eve (thin arrow), who retrieves the information by measuring the photons.}
  \label{fig:1}
\end{figure}

QKD allows two remote parties, usually called Alice (transmitter) and Bob (receiver), to share a common secret key with information theoretical security, over an insecure quantum channel and an authenticated or broadcast classical channel. QKD's security derives from the laws of quantum physics and its implementation necessarily makes use of physical systems, whose correct behavior has to be characterized and guaranteed against unwanted imperfections. Any ignored deviation from the expected behavior can be exploited by an attacker (Eve) to compromise the system security.
In Fig.~\ref{fig:1}, the Trojan-horse attack (THA) against an optical QKD setup is sketched. Eve uses the optical channel connecting Alice and Bob to launch a bright light pulse containing Trojan photons into Alice's supposedly secure module. The light pulse reaches the encoding device and is encoded with the same information $\varphi$ as the photon normally prepared by Alice and then sent to Bob. The information $\varphi$ is meant to be private. However, some of the Trojan photons are reflected back and deliver it to Eve, thus compromising the security of the system.

This eavesdropping strategy was initially described in~\cite{VMH01} and afterwards named ``Trojan-horse attack'' in~\cite{GFK+06}.
Due to its apparent simplicity, the THA has been often considered easily tractable. However, to date, there is no quantitative analysis to mitigate it and there is an increasing number of experiments showing its severity instead~\cite{GFK+06,JAK+14,JSK+14,SRK+14,KJS+14}. For example, it was shown in~\cite{GFK+06} that phase information can be extracted from a LiNbO$_3$-based encoding device using optical-frequency-domain reflectometry. More recently, it has been demonstrated that phase values from an encoding device can be discriminated with 90$\%$ success probability using only 3 photons~\cite{JAK+14}.

To counteract the THA, different solutions have been proposed. On the one hand, active countermeasures, similar to the ones used to ensure the security of the `plug-and-play' QKD setup~\cite{MHH+97,SGG+02,ZQL08,ZQL+10}. Alice could be endowed with an active phase randomizer~\cite{GFK+06,ZQL07} and a watchdog detector~\cite{MHH+97,SGG+02} to remove the phase reference from Eve's hands and bound the energy of the incoming light pulses. However, active components usually add extra complexity to the setup and may offer more options to the eavesdropper~\cite{JSK+14}. For instance, it has been shown recently that a monitoring detector of a commercial QKD system can be bypassed easily~\cite{SRK+14}. On the other hand, passive countermeasures can be realized with much simpler elements, e.g. optical fiber loops, filters and isolators, which leave fewer degrees of freedom to the eavesdropper. Furthermore, they are often inexpensive, simple to implement and to characterize experimentally. However, in this case, powerful resources like the phase randomization and the watchdog detector cannot be used to prove the security of the system.

As a result, the security analysis of the THA remains elusive and no security-proof solution has been derived to date. The only provably secure countermeasures are for users endowed with a teleportation filter~\cite{LC99} or for the receiver in a system running the BB84 protocol~\cite{BB84,VMH01}. In the former case, the solution is not practical and it entails considerable changes in the setup that could open additional loopholes. In the latter case, a delay line installed at the entrance of Bob's module prevents Eve from reading the basis information before the qubit has entered Bob's protected territory. However, the same measure is ineffective to protect the transmitting side of the QKD system, nor does it apply to other protocols such as the B92~\cite{B92} and the SARG04~\cite{SARG04}. Hence it cannot be considered a general solution against the THA.

In this work, we analyze an entirely passive architecture to counteract the THA. We provide quantitative bounds that connect the values of the passive optical components to the security of the QKD system. The key element is interpreting the THA as a side channel. Normally Alice is unaware of it and treats her preparation as ideal. This causes undetected leakage of information from her module to Eve's territory. However, if Alice characterizes the relevant optical components in her apparatus, she can bound the information leakage and attain security through an adequate level of privacy amplification.

\section{Theoretical description}
\label{sec:2}

\begin{figure}[tbp]
%
  \includegraphics[width=8.5cm]{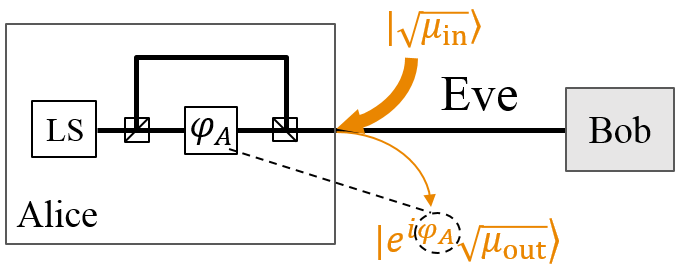}\\
  \caption{Schematics of the transmitting unit of a unidirectional fiber-based QKD setup and Eve's THA. LS is a generic light source. The square with $\varphi_A$ is the encoding device. It writes the phase information $\varphi_A$ on photons traveling in the short arm of the interferometer. Eve injects a bright light pulse in the coherent state $|\sqrt{\mu_{\textrm{in}}}\rangle$ into Alice's module. A fraction of it is encoded by Alice and back-reflected to Eve, emerging as $| e^{i \varphi_A} \sqrt{\mu_{\textrm{out}}}\rangle$, i.e., attenuated by a factor $\gamma$ ($\mu_{\textrm{out}}=\gamma \mu_{\textrm{in}}$) but containing the phase information $\varphi_A$ (dashed line).}
  \label{fig:2}
\end{figure}

\noindent Let us consider the transmitter module~\cite{NOTE1} in the unidirectional, fiber-based, phase-modulated QKD setup depicted in Fig. \ref{fig:2}. In the THA, Eve injects light into Alice's apparatus through the same optical fiber that serves as a quantum channel between the users (thick arrow in the figure). The goal is to reach the phase modulator that encodes the private information $\varphi_A$. A concrete possibility for Eve is to use a laser emitting pulses with average photon number $\mu_{\textrm{in}}$, prepared in a coherent state $|\sqrt{\mu_{\textrm{in}}}\rangle$~\cite{G63}. The pulses acquire the phase modulation information $\varphi_A$ and return to Eve as $ |e^{i \varphi_A} \sqrt{\mu_{\textrm{out}}}\rangle$ (thin arrow in Fig.~\ref{fig:2}), where $\mu_{\textrm{out}}=\gamma \mu_{\textrm{in}}$, being $\gamma \ll 1$ the optical isolation of the transmitting unit. The light pulse retrieved by Eve is correlated to the phase $\varphi_A$ and this compromises the security of the system.

To prevent the THA, Eve's action has to be bounded by a physical mechanism. In particular, it is clear that if the intensity $\mu_{\textrm{in}}$ is unbounded, no solution can exist against the THA. On the contrary, when $\mu_{\textrm{in}}$ is bounded, Alice can adjust the value of the optical isolation $\gamma$ so to make $\mu_{\textrm{out}}$, and therefore Eve's information, arbitrarily small. In this work, we consider the Laser Induced Damage Threshold (LIDT) as the main physical mechanism limiting Eve's action. The LIDT provides an estimate of the energy, thence of the number of photons, that Eve can inject into Alice's module in a characteristic time interval without damaging it. Details about the LIDT are given in Section III. For the moment we call $N$ the maximum number of photons that Eve is allowed to inject in the transmitter module in the time unit (1 second) without violating the LIDT condition. This parameter will be used to provide a
security argument against the THA.

\subsection{Preliminary quantities}
\label{subsec:2A}

\noindent In a THA, Eve firstly prepares $M$ groups of photons and then use each group to probe a different value of Alice's phase modulator (PM). To fix ideas, we can imagine that each group of photons physically corresponds to one pulse of Eve's light source and that each pulse is prepared in a pure coherent state~\cite{P93}. The resulting structure is a tensor product of coherent states:
\begin{equation}
\label{eq:1}
    \ket{\sqrt{\mu_1}}\otimes\ket{\sqrt{\mu_2}}\otimes...\otimes\ket{\sqrt{\mu_M}}.
\end{equation}
In Eq.~\eqref{eq:1}, $\mu_i$ ($i={1,...,M}$) is the mean photon number of the $i$-th coherent state. In order to not overcome the LIDT threshold $N$, Eve has to guarantee the following condition:
\beq
 \sum_{i=1}^M \mu_i = M \mu_{\textrm{in}} < N ,
\eeq
where we have introduced the overall mean photon number of Eve's light $\mu_{\textrm{in}}$.
In general, it is possible for Eve to vary each $\mu_i$ to enhance her strategy. However, it turns out that this brings no advantage to her, as we shall show later on. The convexity of the key rate as a function of $\mu_i$ makes it always better for Eve to set $\mu_{i}$ equal to a constant value. Therefore we have:
\begin{equation}
\label{eq:4}
    \mu_{i} = \mu_{\textrm{in}}.
\end{equation}
It can be noted  that Eq.~\eqref{eq:4} rules out a whole class of attacking strategies by Eve, where she redistributes her initial Trojan photons in a fewer number of pulses. Intuitively, this could increase Eve's information on a subset of Alice's states, but can never increase her total information about the whole key. It is beneficial to Eve to distribute her photons evenly among the available pulses, so to maximise her total information. We will reach the same conclusion in Sec. III.A, but from a physical point of view. There, an even distribution of the Trojan photons will allow Eve to keep the LIDT of an optical component close to its minimum value.

Each of Eve's Trojan pulses is sent in the transmitting unit to probe a different phase value $\varphi_A$ of Alice's PM. After that, the pulses are retrieved by Eve and their mean photon number amounts to $\mu_{\textrm{out}} = \gamma \mu_{\textrm{in}}$. Let us call $f_A$ the total number of phase values encoded by Alice's PM in 1 second. This is equal to the PM clock rate, expressed in Hz. Because Alice knows $f_A$, the maximum number of Trojan photons per second $N$ and the optical isolation $\gamma$, she can bound the mean photon number of the Trojan pulses emerging from her module. We call $\mu_{\textrm{out}}$ the upper bound. It amounts to:
\begin{equation}
\label{eq:3}
    \mu_{\textrm{out}} = \frac{N \gamma}{f_A}.
\end{equation}
$\mu_{\textrm{out}}$ is a crucial parameter in the security argument because it is directly controllable by Alice. It can be interpreted as the mean photon number of the Trojan pulses retrieved by Eve.

In the next section, we proceed from these preliminary observations to derive the secure key rate of the BB84 protocol, assuming that Alice is endowed with an ideal single-photon source. Then, in Section II.C, we extend the security argument to the BB84 protocol implemented with a laser source and decoy states.

\subsection{Key rate of single-photon BB84 protocol}
\label{subsec:2B}

\noindent Let us suppose that Alice prepares ideal single-photon BB84 states and that the only source of information leakage from Alice's system to Eve is from the THA on the PM. Eve shall execute the THA using coherent states of constant intensity as per Eqs.~\eqref{eq:1} and \eqref{eq:4}. We assume the worst-case scenario where Eve can retrieve her states back from the quantum channel with 100$\%$ fidelity, even though, in practice, this may be not fully permitted by the laws of physics.
In this description, the THA can be executed without adding any noise to the communication channel. Despite this, secure keys can still be extracted if the QKD system is well characterized. This is quite counter-intuitive as it challenges the common view of QKD as an \textit{eavesdropping detection} system, while promoting it as an \textit{eavesdropping prevention} system~\cite{SYK14}.

The characterization of the QKD system proceeds as follows. With reference to Alice's interferometer (see Fig.~2), we define the states in the computational basis $Z$ as $\ket{0_Z}=\ket{1}_l\ket{0}_s$, $\ket{1_Z}=\ket{0}_l\ket{1}_s$, where $\ket{n}_l$ ($\ket{n}_s$) is the $n-$photon state traveling in the long (short) arm of the interferometer. Then we write the four BB84 protocol states as $\ket{0_X}$, $\ket{1_X}$ and $\ket{0_Y}$, $\ket{1_Y}$ for the $X$ and $Y$ bases, respectively, corresponding to setting the phase $\varphi_A$ equal to $\{0,\pi\}$ and $\{\pi/2,3\pi/2\}$, respectively, in the qubit
state $(\ket{0_Z}+e^{i \varphi_A}\ket{1_Z})/\sqrt{2}$.

Eve's task is to determine $\varphi_A$ using the light back-reflected from Alice's apparatus. However, the states prepared by Alice and sent to Bob (below labeled with ``B'') are single photons and do not give any phase reference to Eve. Also, the states sent and retrieved by Eve (below labeled with ``E'') originate from an external independent source. Therefore the resulting states emerging from Alice's module can be written as tensor products:
\begin{eqnarray}
\label{eq:7}
\nonumber \ket{\psi_{0X}}_{BE} &=& \ket{0_X}_B \otimes \ket{+\sqrt{\mu_{\textrm{out}}}}_E , \\
\nonumber \ket{\psi_{1X}}_{BE} &=& \ket{1_X}_B \otimes \ket{-\sqrt{\mu_{\textrm{out}}}}_E , \\
\nonumber \ket{\psi_{0Y}}_{BE} &=& \ket{1_Y}_B \otimes \ket{+i\sqrt{\mu_{\textrm{out}}}}_E , \\
          \ket{\psi_{1Y}}_{BE} &=& \ket{0_Y}_B \otimes \ket{-i\sqrt{\mu_{\textrm{out}}}}_E .
\end{eqnarray}
The above states justify an alternative interpretation of $\mu_{\textrm{out}}$, i.e., an excess mean photon number exiting Alice's module. If $\mu_{\textrm{out}}=0$, only true single-photon states leave the transmitting unit, whereas if $\mu_{\textrm{out}}>0$, a hidden side-channel, created by the THA, provides Eve with additional information via the excess photons contained in the states of Eq.~\eqref{eq:7}.

It is natural to ask how Eve can use the information obtained in the THA.
One possibility is for her to wait until the basis reconciliation step of QKD, in order to measure the back-reflected Trojan photons in the correct basis and learn the bit encoded by Alice. In this case, Eve simply prepares and retrieves the Trojan photons and causes no disturbance on the quantum channel. However, she only gains the information carried by the Trojan photons and makes no use of the photons prepared by Alice.
A more powerful strategy is to use the Trojan photons during the quantum transmission, without waiting for the basis reconciliation step. Eve could first measure the Trojan photons and then decide whether to stop or transmit Alice's qubits conditional on the result from her measurement.
Finally, Eve could glean information about the basis chosen by Alice and use it to measure Alice's qubits, thus making optimal use of all the sources of information available to her. This is a convenient framework, as it allows to prove the security of QKD against the most general attack by Eve~\cite{GLLP04,K09,LP07}.
We analyze all the above attacking strategies, from the weakest one to the most general. The first and second THA are analyzed in Appendices~C.1 and~C.2, respectively, while the third, most general, THA is outlined here and detailed in Appendix~B.

To bound the security in the general case, we resort to the so-called ``GLLP approach''~\cite{GLLP04}. More precisely, we use the refinement of GLLP based on the qubit distillation protocol by Koashi~\cite{K09}. In Appendix~B, we apply this approach to the states in Eq.~\eqref{eq:7} and derive the secure key rate of the efficient BB84 protocol~\cite{LCA05,SR08}. There, it is shown that if the key is distilled from the $X$ basis and the phase error rate is estimated in the $Y$ basis, the asymptotic key rate of a QKD system endowed with a single-photon source is:
\beq \label{eq:8}
    R = Q_X [ 1-h(e'_Y) - f_{EC} h(e_X) ].
\eeq
In Eq.~\eqref{eq:8}, $Q_X$ is the single-photon detection rate in the $X$ basis, i.e., the joint probability that a single-photon pulse is emitted by Alice and detected by Bob and both users measure in the $X$ basis; $h$ is the binary entropy function, $f_{EC}$ is the error correction efficiency~\cite{BS93} and $e_X$ is the (single-photon) quantum bit error rate (QBER) measured in the $X$ basis. The term $e'_Y$ is the (single-photon) error rate estimated in a virtual protocol where the users measure in the $Y$ basis and Alice announces the $X$ basis~\cite{K09}. It is given by the following equations:
\bea \label{eq:eY-GLLP}
\nonumber   e'_Y &=& e_Y + 4 \Delta' (1-\Delta') (1-2e_Y)+ \\
\nonumber   & & +4(1-2\Delta')\sqrt{\Delta' (1-\Delta') e_Y (1-e_Y)}~, \\
\nonumber   \Delta' &=& \frac{\Delta}{\mathcal{Y}}~~, \\
            \Delta &=& \frac{1}{2} [1-\exp(-\mu_{\textrm{out}}) \cos(\mu_{\textrm{out}})] ,
\eea
where we conservatively defined $\mathcal{Y}:=\min[\mathcal{Y}_X,\mathcal{Y}_Y]$, with $\mathcal{Y}_X$ and $\mathcal{Y}_Y$ the single-photon yields in the $X$ and $Y$ basis, respectively, i.e., the conditional probabilities that a single photon state emitted by Alice causes a click in Bob's detector, when the users measure in the same basis, $X$ or $Y$.
The presence of $\mu_{\textrm{out}}$ in the last line of Eq.~\eqref{eq:eY-GLLP} shows how the THA affects the key rate in Eq.~\eqref{eq:8}.

The key rate $R$ has been plotted in Fig.~3 as a function of the distance between the users, for different values of the output mean photon number $\mu_{\textrm{out}}$, using parameters close to existing real systems~\cite{LPD+13}.
From the figure, it can be seen that the key rate corresponding to $\mu_{\textrm{out}}=10^{-6}$ is indistinguishable from $\mu_{\textrm{out}}=0$ (no THA) over distances up to 100 km, i.e., about $60\%$ of the maximum distance (170 km in Fig.~3). For $\mu_{\textrm{out}}=10^{-8}$, the key rates in presence and absence of a THA overlap over nearly the whole range. In this case, a negligible amount of additional privacy amplification is required to guarantee security against the THA. The key rate remains positive also for $\mu_{\textrm{out}}=10^{-2}$, but the maximum distance is limited to 9~km in this case and the key rate is severely affected by the THA. The largest value of $\mu_{\textrm{out}}$ showing a positive key rate is $0.015$.
\begin{figure}[tbp]
%
  \includegraphics[width=8.0cm]{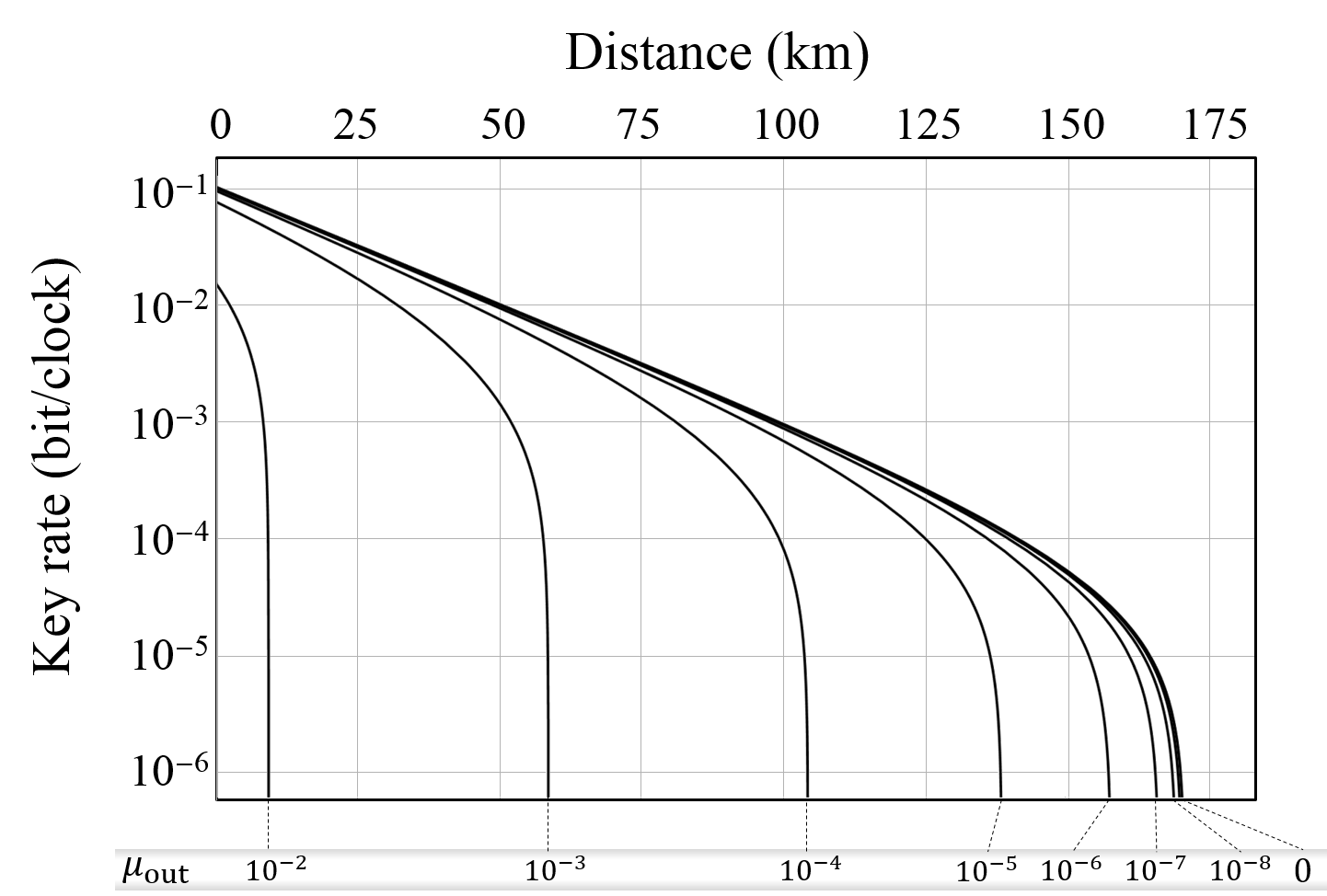}\\
  \caption{Asymptotic key rate $R$ versus distance for the single-photon efficient BB84 protocol under a THA. The rate is plotted for different values of the output mean photon number $\mu_{\textrm{out}}$. Other parameters in the simulation are: fiber loss coefficient $0.2$ dB/km, total detection efficiency $12.5\%$, optical error rate $1\%$, dark count probability per gate $10^{-5}$, error correction inefficiency 20\% above the Shannon limit.}
  \label{fig:3}
\end{figure}

It is worth remarking that the entire effect of the THA is condensed in the parameter $\mu_{\textrm{out}}$, as it is apparent from Eq.~\eqref{eq:eY-GLLP}. Therefore, the obtained key rate is equally applicable to any QKD setup capable of guaranteeing an upper bound to the mean number of Trojan photons reflected by the transmitter back to Eve.

\subsection{Key rate of decoy-state BB84 protocol}
\label{subsec:2C}

\noindent The key rate in Eq.~\eqref{eq:8} has been derived assuming that a single-photon source is available to Alice. However, it is well-known that security can still be guaranteed without a single-photon source if a phase-randomized attenuated laser~\cite{YLD+14} is combined with the decoy-state technique~\cite{W05,LMC05}. Actually such a solution is currently more efficient than a single photon source due to the limited generation rates of existing single photon sources~\cite{BUS+05,NAL+14}.

To extend the result to a decoy-state source, we assume that the decoy-state execution is not affected by the THA. This is equivalent to saying that Eve's only target in the THA considered here is Alice's PM and the devices used by Alice to implement the decoy-state technique are not touched by the THA (see Assumption~3 in Appendix~A and accompanying discussion).
Under this assumption, the decoy-state key rate is a straightforward generalization of Eq.~\eqref{eq:RSP2} along the lines described, e.g., in~\cite{LMC05}. Indicating with a tilde the quantities to be estimated via the decoy-state technique and with $s$ the mean photon number of the signal pulse in the decoy set of states, we obtain:
\begin{equation}
\label{eq:RSP3}
  \widetilde{R} = \widetilde{Q}_X^{(1)}  \left\{ 1-h \left[ \tilde{e}^{\prime(1)}_Y \right] \right\} - Q_X^{(s)} f_{EC} h[e_X^{(s)}],
\end{equation}
where:
\bea \label{eq:25tris}
\nonumber   \tilde{e}_Y^{\prime (1)} &=& \tilde{e}_Y + 4 \widetilde{\Delta}' (1-\widetilde{\Delta}') (1-2 \tilde{e}_Y)+ \\
\nonumber                &+& 4(1-2 \widetilde{\Delta}')\sqrt{\widetilde{\Delta}' (1-\widetilde{\Delta}') \tilde{e}_Y (1-\tilde{e}_Y)},\\
            \widetilde{\Delta}' &=& \frac{\Delta}{\widetilde{\mathcal{Y}}}.
\eea
In Eq.~\eqref{eq:RSP3}, $\widetilde{Q}_X^{(1)}$ is the decoy-state estimation of the single-photon detection rate $Q_X$ in Eq.~\eqref{eq:8}, while $Q_X^{(s)}$ is the detection rate of the signal pulse measured in the $X$ basis.
In Eq.~\eqref{eq:25tris}, we conservatively defined $\widetilde{\mathcal{Y}}=\min[\widetilde{\mathcal{Y}}_X,\widetilde{\mathcal{Y}}_Y]$, with $\widetilde{\mathcal{Y}}_X$  and $\widetilde{\mathcal{Y}}_Y$ the single-photon yields in the $X$ and $Y$ basis, respectively, estimated via the decoy state technique.

\begin{figure}[tbp]
%
  \includegraphics[width=8.5cm]{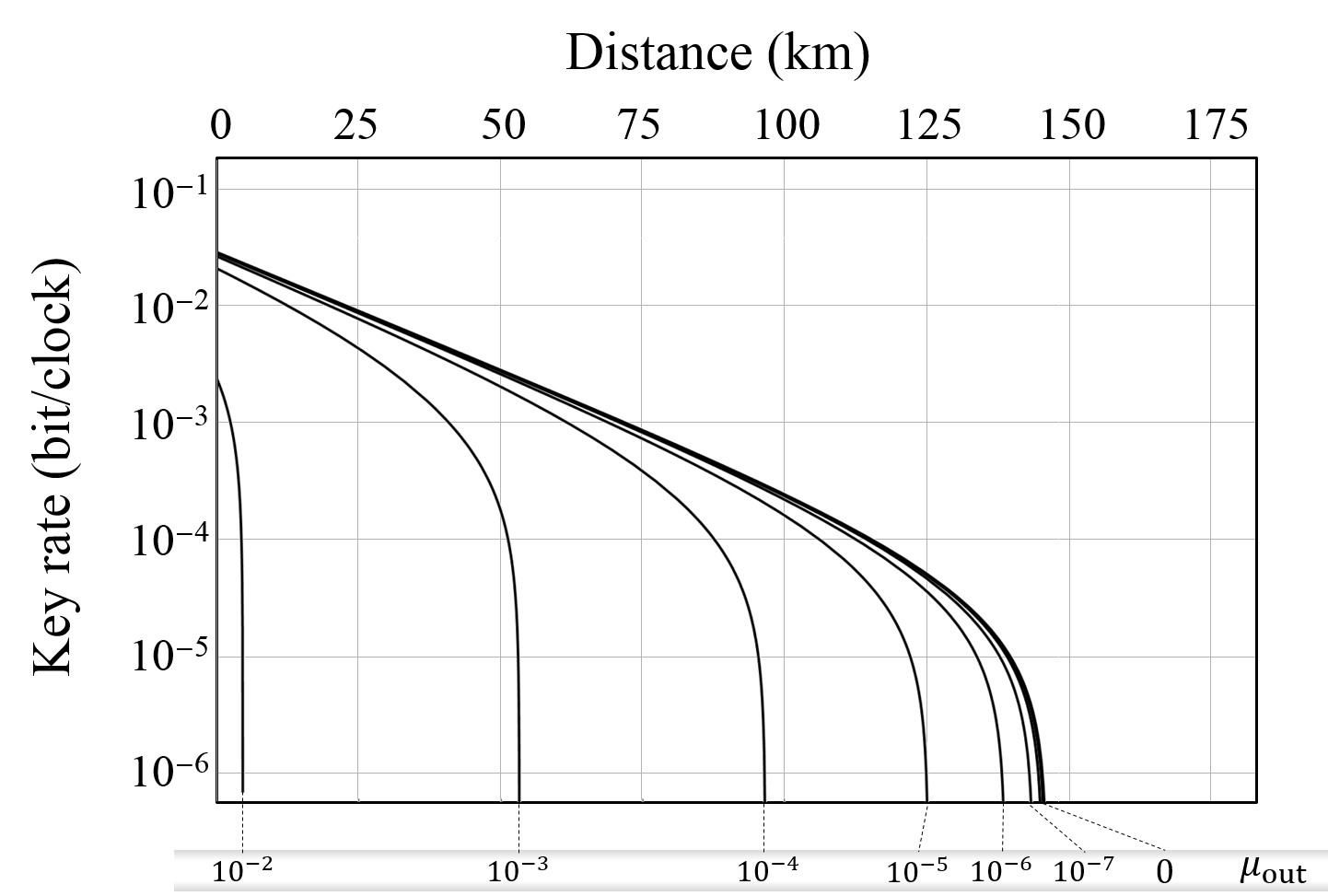}\\
  \caption{Asymptotic key rate $\widetilde{R}$ versus distance for the decoy-state efficient BB84 protocol under a THA, for various values of the output mean photon number $\mu_{\textrm{out}}$. Experimental parameters in the simulation are as in Figs.~\ref{fig:3}. The average photon number of the signal states in the decoy-state technique is $s=0.5$.}
  \label{fig:4}
\end{figure}

The key rate $\widetilde{R}$ is plotted in Fig.~\ref{fig:4}. Although rate and maximum distance are smaller than in the single-photon case (Fig.~\ref{fig:3}), as expected, it is remarkable that the key rate corresponding to a value $\mu_{\textrm{out}}=10^{-7}$ remains indistinguishable from the ideal rate ($\mu_{\textrm{out}}=0$) over nearly the whole distance range. A 10 times larger value, $\mu_{\textrm{out}}=10^{-6}$, which is easier to achieve in practice, generates a key rate that closely follows the ideal one up to 100~km, i.e., $70\%$ of the maximum distance achievable (146~km in Fig.~\ref{fig:4}), and remains positive up to 140~km, i.e., 96\% of the maximum distance. This motivates our choice of $\mu_{\textrm{out}}=10^{-6}$ for the case study in Sec.~\ref{sec:4}.A. Finally, it is worth noting that the key rate remains positive even for larger values of $\mu_{\textrm{out}}$, up to $0.012$.

Before concluding this section, a couple of remarks are in order. First, it has been convincingly proven that the key rate achieved with only three decoy states is very close to that obtained with an infinite amount of decoy states~\cite{MQZ+05}. We have run simulations that confirm this result. Therefore the key rate in Fig.~\ref{fig:4} is achievable in a real system.
Second, the rate equations provided so far have been derived using coherent states of constant intensity. Here we show that this setting is actually advantageous to Eve.
Suppose that Eq.~\eqref{eq:4} does not hold, i.e., $\mu_i\neq\mu_{\textrm{in}}$. With this setting, Eve is trying to distribute her $N$ Trojan photons unevenly among the $M$ pulses, in an attempt to enhance her information gain. Suppose that Eve distributes the $N$ photons in only two classes of pulses, such that the first (second) class features an average photon number $\mu_1$ ($\mu_2$) and $\mu_1 < \mu_2$ , $\mu_{\textrm{in}}=(\mu_1+\mu_2 )/2$. Then, for each of the key rates given in this work, represented by a generic symbol $\mathcal{R}$, we have numerically verified that:
\begin{equation}
\label{eq:11}
    \mathcal{R}(\mu_{\textrm{in}}) \leq \frac{\mathcal{R}(\mu_1)+\mathcal{R}(\mu_2)}{2}.
\end{equation}
For that, we have used the explicit expressions of the key rates and their dependance on $\mu_{\textrm{out}}$, which is related to the input photon number by the linear equation $\mu_{\textrm{out}}=\gamma\mu_{\textrm{in}}$. In other terms, the key rates are convex functions of $\mu_{\textrm{out}}$ and thence of $\mu_{\textrm{in}}$. According to Eq.~\eqref{eq:11}, the rate distilled by the users under Eve's new strategy (R.H.S.) is larger than the one pertaining to the old strategy (L.H.S.), so the new strategy is less effective and not advantageous to Eve.
More general strategies by Eve that account for more than two classes of photons with different mean photon numbers can be treated as a trivial extension of Eq.~\eqref{eq:11}.

\section{Bounds on input photons}
\label{sec:3}

\noindent In our security argument, the quantity $N$ plays an important role. Here we provide more details about this limiting threshold and describe a way to quantify it. For that, we adopt a pragmatic approach, motivated by the results of the
previous sections. In particular, we have established that Eve conducts the THA using coherent states of constant intensity. Therefore we can conveniently think that such states are generated by a single-mode laser operated well above threshold~\cite{L73}. This view naturally leads to considering the laser-induced damage threshold, LIDT. From a security perspective, the LIDT can only provide a general indication of the bounds to be used in a security analysis. The actual response to thermal damage of the real components of a QKD system should be experimentally measured.

\subsection{Laser-induced damage threshold}
\label{subsec:3A}

\noindent A single-mode optical fiber is arguably the most common component of a fiber-based QKD setup. It is used mainly to transmit information in the third telecom window (wavelength $\lambda=1.55~\mu$m) because of its small attenuation coefficient. Its typical core diameter is $8-10~\mu$m, corresponding to a core area of $50-80~\mu$m$^2$. If the laser power used by Eve is sufficiently high, it creates an accumulation of energy in this small region of the core and increases the temperature of the medium beyond the tolerance level, inducing fiber thermal damage~\cite{AGM+85,TB93}.

Such a damage threshold is usually quantified by the LIDT, laser-induced damage threshold, defined in the 2011 international standard ISO 21254-1 as \cite{ISO}: ``\textit{the highest quantity of laser radiation incident upon the optical component for which the extrapolated probability of damage is zero, where the quantity of laser radiation may be expressed in energy density, power density or linear power density}'' \cite{NOTE2}. The smaller the LIDT of the component, the larger the probability to damage it. This subject is well-studied and values for the LIDT of a silica-based optical fiber, which is the component we are interested in, can be obtained. However, before discussing the absolute values, it is worth examining the qualitative behavior of the LIDT, which is determined by the underlying thermal damaging mechanism. The purpose is to investigate how features of Eve's laser like the repetition rate or the pulse width can affect the LIDT and, by consequence, the system security. This provides useful indications for setting a proper LIDT value.

One prominent feature of the LIDT is that it increases with the pulse width of the incident laser, i.e., a wide light pulse causes less damage to the optical component than a narrow one. This makes narrow pulses more detectable to Alice and Bob than wide ones. This can be formalized using the well-known square root dependence of the LIDT on the pulse width~\cite{W03,DLK+94,SFR+95,MLR+05}:
\begin{equation}
\label{eq:13}
    \frac{\textrm{LIDT}(\tau_1)}{\textrm{LIDT}(\tau_2)} = \sqrt{\frac{\tau_1}{\tau_2}} .
\end{equation}
Here $\tau_1$  and $\tau_2$ are two different pulse widths for the same pulse energy. Eq.~\eqref{eq:13} suggests that Eve's laser pulse should be the widest possible, compatible with Alice's phase modulator.

A similar rule applies to the laser wavelength, resulting in the shorter wavelength to be causing more damage to the optical component than the longer one (see e.g.~\cite{CRD03}):
\begin{equation}
\label{eq:14}
    \frac{\textrm{LIDT}(\lambda_1)}{\textrm{LIDT}(\lambda_2)} = \sqrt{\frac{\lambda_1}{\lambda_2}} .
\end{equation}
Eq.~\eqref{eq:14} suggests that Eve's optimal laser's wavelength should be as large as possible, even larger, if necessary, than the typical wavelength used in the QKD setup. However, it also entails that the LIDT remains reasonably constant for all the wavelengths possibly transmitted in the fiber. A standard optical fiber cannot transmit by total internal reflection beyond the so-called ``bend-edge'' wavelength, which is only a few hundred of nanometers away from the fiber cutoff wavelength (see, e.g.,~\cite{Fibercore}). As an example, we can consider a bend-edge wavelength of $1850$~nm for an optical fiber transmitting at $1550$~nm~\cite{WFF05,NOTE7}. According to Eq.~\eqref{eq:14}, this would increase the LIDT by less than $10\%$, showing that the wavelength of Eve's laser is not crucial in determining the efficacy of the THA. To compensate for this effect in the theory, it suffices to increase the LIDT value by 10\%.

To upper bound the input photon number $N$ used in the security argument, we need to estimate the LIDT of Alice's optical module. This is arguably given by the LIDT of the most fragile component in the module. However, we will consider the LIDT of just one of the components in Alice's unit, the one most exposed to Eve's light. The other components are assumed to either work in their normal operation regime or fail in a way that is detectable by the users (see Assumption~1 in Appendix~A and accompanying discussion). In Section IV, we describe the architecture of Alice's setup against the THA. The component most exposed to Eve's light is a loop of standard optical fiber placed at the main entrance of the transmitting box. Hence we are interested in the LIDT of a standard single mode optical fiber. One possible way to estimate it, is to consider the geometry of the fiber and the material it is made of. As said, a typical fiber has a core area of about $50~\mu$m$^2$ and is made of fused silica. The LIDT of fused silica is determined by the softening point of the material~\cite{W03} and amounts to $1.1\times10^7$~J/cm$^2$~\cite{NOTE3}. For a longer time, the silica-based medium starts dissipating heat and the threshold increases linearly with the pulse width. For a shorter time, the square root law in Eq.~\eqref{eq:13} applies, decreasing the LIDT accordingly.

The above-cited LIDT value corresponds to an average power of $5.5 \times 10^4$~W over $50~\mu$m$^2$. For a typical wavelength of $\lambda=1.55~\mu$m this means that $4.3 \times 10^{23}$ photons impinge every second onto the fiber core area. Before such a large number of photons can damage the fiber core, other highly detectable damages are likely to occur at the fiber interfaces, causing, e.g., a net reduction of the transmission, or an increase in the noise figure. Also, the LIDT value mentioned above relates to a homogeneous medium. In reality, large temperature gradients can occur in the proximity of a defect, or at the connection between two segments of fiber, or at the interface between the fiber core and the cladding. Some of these properties can even be artificially enhanced by acting on the number of connectors, the bending radius and the doping levels of the fiber. These considerations lead to the conclusion that the given LIDT value is an overly conservative estimation of the real LIDT of an optical fiber. In the next section, we obtain a different LIDT value by combining the findings of Section~II with the results from experiments performed on real optical fibers.

\subsection{Fiber thermal fuse-induced LIDT}
\label{subsec:3B}

\noindent In Section~II, we have shown from an information theory point of view that Eve's optimal strategy is to distribute her photons into a number of pulses $M$ that is equal to Alice's PM clock rate (in Hz) $f_A$, so to maximize her total information gain. In the description of Eve's laser, the above strategy translates into setting $f_E = f_A$, where $f_E$ is Eve's laser repetition rate. Moreover, in the previous section, we have shown that the LIDT depends only weakly on the laser pulse width and that the larger the width, the larger the damaging threshold. In the description of Eve's laser, this translates into having a laser pulse width, $\tau_E$, as large as possible, compatibly with Alice's PM. Let us call $\tau_A$ the time window of Alice's PM. If $\tau_E>\tau_A$, a fraction of Eve's photons falls outside the PM gate and deliver no information to Eve. Therefore, the optimality condition for Eve is $\tau_E=\tau_A$. This condition on the pulse width represents an additional constraint for Eve and an extra parameter under Alice's control. After $\gamma$ and $f_A$, Alice can now act on $\tau_A$ to make Eve's strategy less effective. In particular, by reducing $\tau_A$, Alice reduces the damaging threshold of her module, thence $N$.

Let us draw a worst-case scenario from the above considerations. We conservatively assume that Alice's PM is driven by a perfectly rectangle wave. This helps Eve matching the condition $\tau_E=\tau_A$ and simultaneously keeping the damaging threshold high. As a consequence, the amplitude of Alice's PM is assumed to be flat in time. The amplitude is selected at random among the four equally spaced values of the BB84 protocol. The way these values are selected depends on the logic driving the PM. If a non-return-to-zero (NRZ) logic is used, the PM duty cycle is $100\%$, i.e., the PM is always active, transiting from a given phase value directly to the next one and we have in this case $\tau_A=1/f_A$. If a return-to-zero (RZ) logic is used, the modulator is reset after each encoded phase value. In this case the duty cycle is less than $100\%$ and the PM time duration is $\tau_A<1/f_A$. We note that in the particular case Alice's PM is driven according to a NRZ logic ($100\%$ duty cycle), Eve's laser coincides with a continuous-wave (CW) laser, as it emits a seamless sequence of rectangle pulses, all of the same amplitude, sitting one next to each other. A deeper thought reveals that this is actually a worst-case scenario, because when the condition $\tau_E=\tau_A$ is matched, $\tau_E$ takes on its maximum value ($1/f_A$), thus minimizing the risk of optical damage, while leaving Eve's information unchanged. Because of that, we can always imagine that Alice's PM is driven by a NRZ logic, even if it is RZ and, accordingly, that Eve uses a CW laser to probe the PM.

Experiments performed with CW lasers on real optical fibers have demonstrated that an average power around 2-5~W cause a catastrophic thermal damage in a standard single-mode silica fiber~\cite{K88,KB88,NOTE4}. This effect is
known as ``self-propelled self-focusing'' or ``fiber thermal fuse''~\cite{K88,KB88,DMD96,YAN02,ARD+10,TI04,K13,Tod15}. The high power of the laser generates a heating point in the fiber where the local temperature overcomes the melting point of the medium. From there, the damage propagates along the fiber, eventually making it unusable.
This effect has also been exploited to build an ``optical fuse'' that breaks by 1.2-5.3~W of incident light at wavelengths around 1500~nm~\cite{TI04}.
For a wavelength of $\lambda=1550$~nm, $2$~W correspond to $1.6 \times 10^{19}$ photons crossing every second a $50~\mu$m$^2$-fiber core area (a$_{50}$). In order to have an easy reference for the LIDT value, we set it equal to $N=10^{20}$ photons/s/a$_{50}$. The new LIDT value is $4.3\times10^3$ smaller than the previous one. Still, it corresponds to $12.8~W$ from a CW laser, which is much larger than the power threshold reported in the fiber thermal fuse experiments.
We will adopt this number to draw an example where the values of the optical components in Alice's apparatus are connected to the security requirements. However the more conservative threshold for $N$ could be adopted instead to arrange a different use case for an application that requires a stronger bound, independent of the fabrication details of the fiber and relying only the softening point of silica.

\section{Experimental characterization}
\label{sec:4}

\subsection{Passive architecture against the THA}
\label{subsec:4A}

\begin{figure}[tbp]
%
  \includegraphics[width=8.5cm]{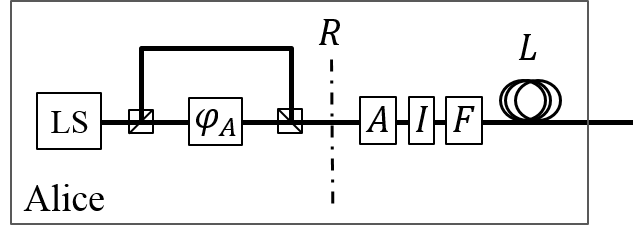}\\
  \caption{Architecture of a QKD transmitter to mitigate the THA. $OFL$: optical fiber loop determining the LIDT; $F$: optical filter; $I$: optical isolator; $A$: attenuator; $R$: total reflection from all components to the left of the dot-dashed line.}
  \label{fig:6}
\end{figure}

\noindent An entirely passive architecture against the THA is drawn schematically in Fig.~\ref{fig:6}.
It is based on a sequence of components that actualize the security argument described so far. A silica-based optical fiber loop (OFL) of length $L$ defines the LIDT of the transmitter and is followed by a filtering block $F$, an optical isolator $I$ and an attenuator $A$. We also indicate with $R$ the total reflectivity of the optical elements to the left of the dot-dashed line (not to be confused with the key rate $R$ given in Eq.~\eqref{eq:8} and plotted in Fig.~\ref{fig:3}). The line for the reflectivity $R$ is conservatively drawn to include also the first beam splitter as seen by Eve to allow an easier experimental implementation (Section IV.B). In the figure, all the components are presumed to either work as expected or fail in a way that is detectable by the users.

The OFL acts as a regulator for high-power input light and as a filter for wavelengths longer than the bend-edge point. Together with the optical filter $F$, which is tuned to let pass the wavelength of the quantum channel and stop all the others, it limits the maximum number $N$ of photons that Eve can inject into Alice's module in the chosen time unit. In other terms, it represents the optical component to which the LIDT should apply. To fix the ideas, we can imagine a length for the OFL greater than $1~$m, in line with what was reported in the experiments about the fiber thermal fuse effect. A longer OFL can only be beneficial to the users, as it increases the probability of thermal damage, which increases as well if a few interfaces are present in the OFL.

The optical isolator strongly attenuates the input light from Eve, enforcing the unidirectionality condition in the module. A typical dual-stage optical isolator features an isolation value of $10^{-5}$ or smaller. It is convenient to measure the isolation in decibel, or dB, rather than in absolute value. If $x$ is the absolute value of the optical isolation of a given component, we use the following notation
\begin{equation}
\label{eq:15}
    \dot{x} = 10 \log_{10} x
\end{equation}
to indicate its value in decibel. So for example, the optical isolator mentioned above would feature an isolation of $-50$~dB.

The attenuator box in Fig.~\ref{fig:6} is already present in the schematics of various QKD systems using an attenuated laser as light source, while it is not present in systems using a single-photon source, as it would entail major losses in the system. If used, it helps to avert the THA, as it contributes to the optical isolation of Alice's module, $\gamma$. The following equation quantifies the contributions of each single conceptual block in Fig.~\ref{fig:6} to $\gamma$:
\begin{equation}
\label{eq:16}
    \gamma = F^2 \times I^n \times A^2 \times R .
\end{equation}
Eq.~\eqref{eq:16} can be conveniently rewritten in dB:
\begin{equation}
\label{eq:17}
    \dot{\gamma} = 2 \dot{F} + n\dot{I} + 2 \dot{A} + \dot{R}.
\end{equation}
In Eqs.~\eqref{eq:16}, \eqref{eq:17}, the typical double-pass of a THA through Alice's components has been considered. This leads to explicit corrections for the filter and the attenuator terms. For the isolator term there is no such correction, because one direction of the double-pass features zero attenuation. However, there is a factor $n$ that represents the number of optical isolators present in the system.

To relate the isolation $\gamma$ to the system security, we need to connect it to the parameter $\mu_{\textrm{out}}$ via Eq.~\eqref{eq:3}. For that, we introduce the dimensionless ratio $\chi:=N/f_A$ and rewrite Eq.~\eqref{eq:3} in dB notation:
\begin{equation}
\label{eq:18}
    \dot{\mu}_{\textrm{out}} = \dot{\chi} + \dot{\gamma}~.
\end{equation}
To give an example of how Eqs.~\eqref{eq:17} and~\eqref{eq:18} can be used to meet the security criterion, let us start from setting a target value for the excess average photon number $\mu_{\textrm{out}}$. We have seen from Figs.~\ref{fig:3}
and~\ref{fig:4} that a value $\mu_{\textrm{out}}=10^{-6}$ (i.e. $\dot{\mu}_{\textrm{out}}=-60$~dB) can guarantee security against the THA with only a negligible (limited) amount of additional privacy amplification over short-range and middle-range (long-range) QKD transmissions. Therefore we choose this value as the target.
We consider the threshold value $N=10^{20}$ photons/s/a$_{50}$ discussed in Section III.B and a system clock rate $f_A=10^9$~Hz. This gives $\chi=10^{11}$ ($\dot{\chi}=110$~dB).
From Eq.~\eqref{eq:18}, we then get: $\dot{\gamma}=\dot{\mu}_{\textrm{out}}-\dot{\chi} = (-60-110)$~dB $=-170$~dB. This is the total optical isolation required in Alice's module in order to guarantee security.
Alice can try and match this value using well-characterized components and then applying Eq.~\eqref{eq:17}.

\begin{table}
%
%
\begin{center}
\begin{tabular}[c]{|c|c|c|c|c|c|} \hline
%
~Clock rate~ & ~~$f_A$ (Hz)~~ & ~~~$|\dot{\gamma}|$~~~ &~~$|\dot{R}|$~~ & ~~$|\dot{A}|$~~ & ~$|\dot{I}|(n)$~ \\
\hline ~~~1GHz~~~ & ~~$10^9$~~ & ~~170~~ & ~~40~~ & ~~35~~ & ~~60(1)~~ \\
\hline ~~~1GHz*~~ & ~~$10^9$~~ & ~~170~~ & ~~50~~ & ~~~0~~ & ~~60(2)~~ \\
\hline ~~~1MHz~~~ & ~~$10^6$~~ & ~~200~~ & ~~40~~ & ~~30~~ & ~~50(2)~~ \\
\hline ~~~1MHz*~~ & ~~$10^6$~~ & ~~200~~ & ~~50~~ & ~~~0~~ & ~~50(3)~~ \\
\hline ~~~1kHz~~~ & ~~$10^3$~~ & ~~230~~ & ~~40~~ & ~~35~~ & ~~60(2)~~ \\
\hline ~~~1kHz*~~ & ~~$10^3$~~ & ~~230~~ & ~~50~~ & ~~~0~~ & ~~60(3)~~ \\
\hline
\end{tabular}
\caption{\label{tab:1} Practical combinations of system components to meet the target $\mu_{\textrm{out}}=10^{-6}$ when $N=10^{20}$~photons/s/a$_{50}$ and $\dot{F}=0$~dB. All dotted quantities are in decibel and are given in absolute value. Lines with the asterisk are cases in which attenuation cannot be used, e.g., when the transmitter uses a single-photon source or at the receiver side. The feasibility of the values for 1~GHz clock rate has been confirmed experimentally using the QKD setup described in~\cite{DYD+15} .}
\end{center}
\end{table}

Table~\ref{tab:1} contains some possible combinations of $f_A$, $\dot{R}$, $\dot{A}$ and $\dot{I}$ to match the target value $\dot{\mu}_{\textrm{out}}=-60$~dB. For convenience, we report the absolute values of the components. In the table, we set $\dot{F}=0$, for the filter insertion loss is typically close to zero at its central wavelength, and we assume that the filter is centered at the operational wavelength of the QKD setup.
In the first column, we have considered three interesting and feasible regimes, 1~kHz, 1~MHz and 1~GHz. The lines with the asterisk are for situations where attenuation cannot be used, e.g., if the transmitter uses a single-photon source or at the receiver side. It is worth noting that single-photon sources up to the MHz range are currently available (see, e.g., \cite{BUS+05,NAL+14}). In all cases, we have reported what we believe the most practical combination of components could be.
For the optical reflectivity $\dot{R}$, we have considered a typical absolute value of $40$~dB, which comes from a common fiber connector. However, an absolute value of $50$~dB is possible if angled connectors or splicing are used for the fiber-integrated optics in the module. This latter option is worth considering especially for the lines with the asterisk in Table~\ref{tab:1}.
For the optical isolator, its absolute value is set in the factory and cannot be varied by the users. We set it equal either to $50$~dB or to $60$~dB in Table I, according to the most convenient configuration. The former value is common in dual-stage optical isolators. The latter value is less common,
but it can be obtained by properly sampling a set of isolators and selecting the best one (see Sec.~IV.B).
Finally, for the attenuator, we avoided using absolute values larger than $35$~dB, as that would commit the transmitter to unusually high-power lasers.

From Table~\ref{tab:1}, it can be seen that two or more optical isolators might be necessary to meet the security target $\mu_{\textrm{out}}=10^{-6}$. However, if the clock rate is high enough, a single isolator is sufficient (first line of the table). In any case, given the low cost and the low insertion loss of filters, attenuators and optical isolators, all the options in Table I can be considered feasible and relatively inexpensive.

\subsection{Components characterization}
\label{subsec:4B}

\noindent To prove the attainability of the values reported in the first line of Table~\ref{tab:1}, we have experimentally characterized reflectivity and transmission of the components most relevant to security in the transmitting unit of a unidirectional GHz-clocked QKD system~\cite{DYD+15}, within their operational range. A full-range characterization of the real components in the setup is necessary to guarantee their behavior against unwanted deviations, as required by the security argument (see Assumption~1 in Appendix~A).
\begin{figure}[tbp]
%
  \includegraphics[width=8.5cm]{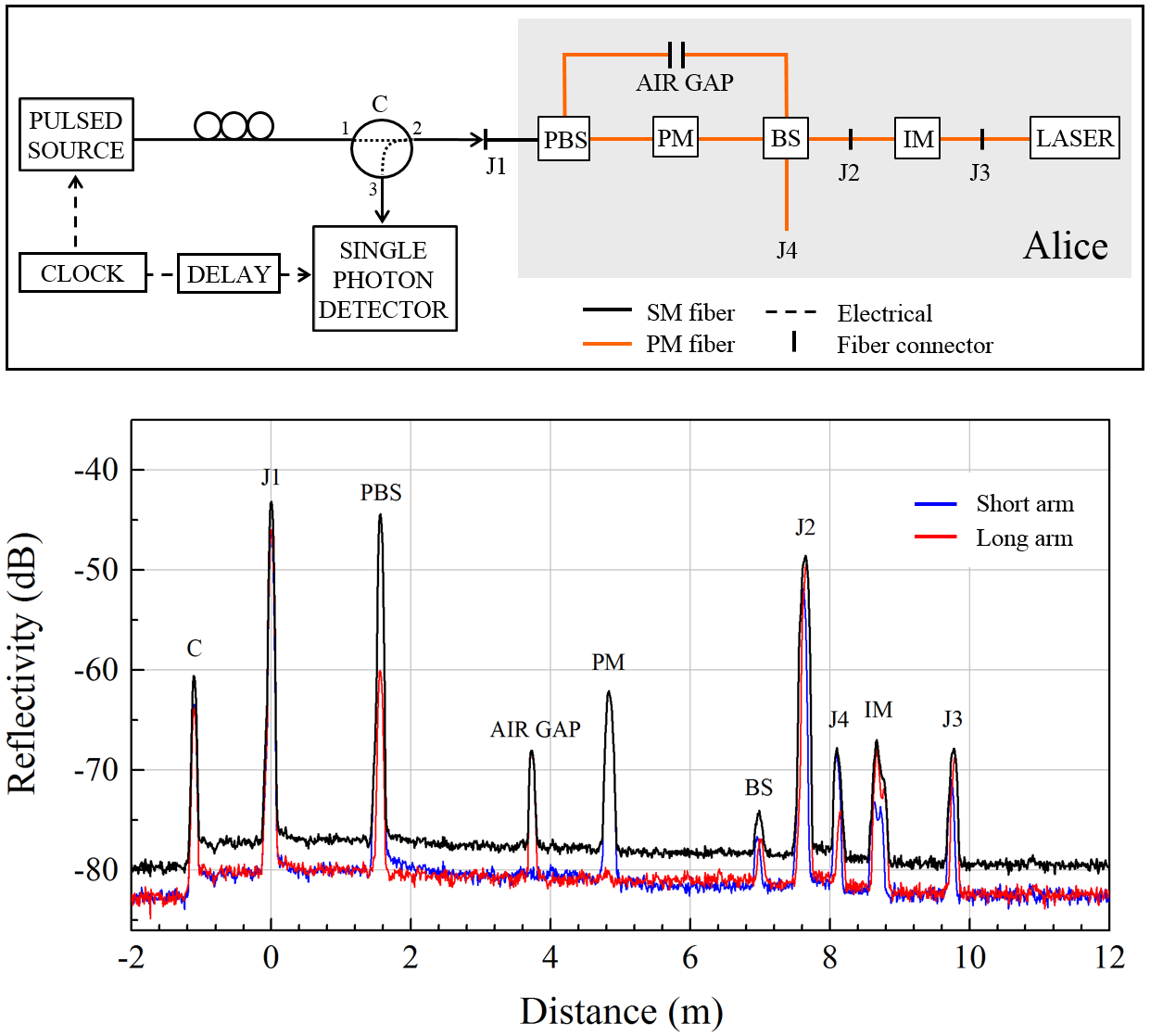}\\
  \caption{\textit{Top} - Schematics of the QKD transmitter module and of the $\nu$-OTDR setup used
  for characterizing its reflectivity. \textit{Bottom} - Reflection peaks of the transmitting unit.
  The distance is measured from the connector J1 placed at the entrance of the module. The traces are
  acquired for two orthogonal polarizations, aligned to maximize the transmission through the short
  (blue traces) and long (red traces) arm of the interferometer. The peaks of the reflectivity are added
  to obtain a worst-case estimation (black). Only the peaks from the components included in the shade
  region of the top diagram have to be considered in the estimation of $R$.}
  \label{fig:7}
\end{figure}

As a first step, we have used single-photon optical time-domain reflectometry ($\nu$-OTDR,~\cite{ELZ+10}) to quantify the reflectivity $R$ of Alice's apparatus. The measurement setup and the resulting traces are shown in Fig.~\ref{fig:7}, on the top and bottom diagrams, respectively. In the $\nu$-OTDR setup, a $1$-MHz pulsed laser at 1550~nm is connected to Alice via a circulator. Polarization controllers are used to align the pulses to the long or short path of Alice's interferometer, so to obtain the output patterns of the orthogonal polarizations. These are shown as blue and red traces in Fig.~\ref{fig:7}. The two patterns have been added together to upper bound the total reflectivity and this is indicated by the black trace in the figure. The upper bound to $R$ is obtained assuming the linearity of the reflectivity, as follows: $R(a|s\rangle + b|l\rangle)=aR(|s\rangle) + bR(|l\rangle)\leq R(|s\rangle)+R(|l\rangle)$, where the vector $|s\rangle$ ($|l\rangle$) represents the polarization traveling in the short (long) arm and $a$, $b$ are complex numbers with modulo squared adding to $1$. The traces are plotted from the entering point of Alice's module, which is connector J1 in Fig.~\ref{fig:7}. However, only the peaks pertaining to the components included in the shaded region of the top diagram have to be considered in the estimation of $R$ (see also dot-dashed line in Fig.~\ref{fig:6}).

The sum of all the peaks relevant to $R$ gives a total reflectivity of $-42.87$~dB. This value meets the requirement $\dot{R}<-40$~dB set in the first line of Table I. Also, the characterized QKD system includes an attenuator set to $-35$~dB.
\begin{figure}[tbp]
%
  \includegraphics[width=8.5cm]{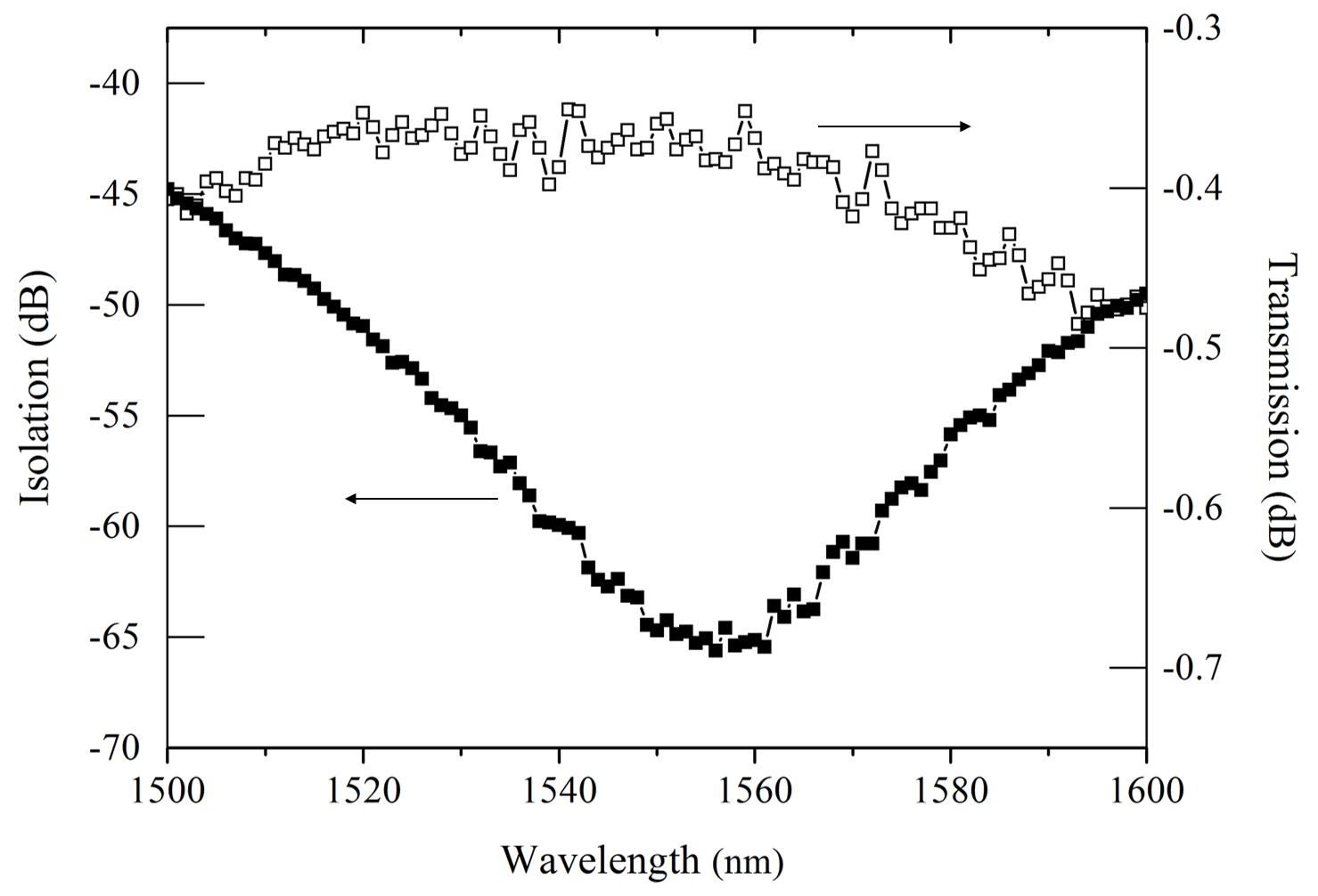}\\
  \caption{Spectral characterization of a dual-stage optical isolator. The isolator shows less than 0.36~dB insertion loss in the forward direction (right axis, empty squares) and more than 65~dB isolation in the backward direction (left axis, filled squares) around the central wavelength of 1550~nm.}
  \label{fig:8}
\end{figure}
To match the $|\dot{\gamma}|=170$~dB condition, additional optical isolation of at least $-60$~dB is needed. Dual-stage isolators specifying typical isolation at this level are commercially available from a number of manufacturers. One isolator was directly tested by us and featured an absolute isolation larger than $65$~dB in the proximity of the main transmission wavelength of the system, $1550$~nm, as shown in Fig.~\ref{fig:8}. Across the S-, C- and L-Band, the isolation value varies, until it reaches a minimum of about $40$~dB. However, in this regime, the optical filter takes over and provides high optical isolation so that a typical suppression of more than $80$~dB is obtainable across the entire C-band. Because the filter is crossed twice by Eve's light, this leads to more than $160$~dB additional optical isolation to the system whenever the wavelength is different from 1550~nm. This demonstrates that the values reported in the first line of Table I are feasible when devices are operated in their working regime.

\section{Discussion}
\label{sec:5}

\noindent In the first part of the work, we derived our main result, i.e., the secure key rate of a QKD system in the presence of a THA, under reasonable assumptions (see Appendix~A for a summary and a discussion of the assumptions). The result depends on Alice's ability to limit the number of the incoming photons $N$ and to reliably upper bound the mean number of Trojan photons $\mu_{\textrm{out}}$ exiting from her module. However, the curves plotted in Figs.~\ref{fig:3} and~\ref{fig:4} are independent of $N$ and can be applied to different QKD systems, provided that the assumptions in the theory are met. From the key rates, we have shown that a value of the mean output photon number $\mu_{\textrm{out}}\sim10^{-6}$ allows to approach the situation with no THA for nearly any distance between the users. For distances up to 70\% of the maximum working distance, this can be achieved without any additional privacy amplification.

In Section III, we have drawn an example of how to set a value on $N$ using the thermal damage point of an optical fiber. The most conservative value for $N$ is in the order of $10^{23}$ photons injected every second on the core area of the fiber. This value has been obtained from the softening point of a homogeneous medium made of fused silica and is independent of future advancements in technology, provided that the composition of fused silica remains unchanged.
A lower value for $N$, equal to $10^{20}$ photons/s/a$_{50}$, has been drawn from recent experiments on the thermal damage of real fibers, after taking into account the presence of inhomogeneities in the fiber and the qualitative behavior of the LIDT in response to the laser pulse width. Using this lower value, we showed the feasibility of our passive architecture in a practical scenario. We related the key rate of a QKD system to its clock, detection rate, reflectivity and to the properties of a sequence of fiber loop, filter and optical isolator, as depicted in Fig.~\ref{fig:6}.
In Table~I, we devised various combinations of these components to meet the security condition against the THA. According to the table, most of the existing QKD systems can potentially be protected from THA's, provided that a sufficient number of optical isolators are used and that the real components behave as expected.

Some elements in our security argument may appear optimistic, for instance, the use of coherent states by Eve. However, we believe that our analysis is overall conservative. The considered LIDT threshold corresponds to a light power of 12.8~W from a CW laser and is larger than the power required to activate the fiber thermal fuse effect in a standard single-mode fiber. It is reasonable to think that before this large number of photons can melt the fiber core, some other mechanism would make Eve detectable. We also assumed a noise-free retrieval of quantum states by Eve, while it is well known that the retrieval is physically limited by Raman and Rayleigh scattering. We decided not to consider the fact that other QKD components, already present in Alice's module, could have a lower LIDT than the optical fiber. Finally, we ignored that monitoring detectors are already present in most of the QKD systems, mainly for stabilization purposes. Such devices additionally constrain Eve's action and can be beneficial to improve our solution.

\section{Conclusion}
\label{sec:6}

\noindent In this work, we studied the security of a fiber-based QKD setup endowed with passive optical components against the long-standing Trojan-horse attack (THA). In the framework of Ref.~\cite{GLLP04}, we provided quantitative security bounds, easily applicable in practice, against a general THA. With the proof method of Ref.~\cite{TCK+14}, we analyzed two specific examples of a THA, giving useful insights into the THA mechanism and the method employed to prove security against it (Appendix~C).
In our analysis, we focused on a particular unidirectional QKD setup, in which light flows from the transmitter to the receiver and the reverse direction is forbidden.
This architecture is similar to that of the transmitters used in Ref.~\cite{LCQ12} to guarantee the measurement-device-independent security of the decoy-state BB84 protocol. Hence we expect that our results can be applied to that system after minor modifications.
The unidirectional configuration allows the use of optical isolators, whose proper behavior has to be tested against undesired deviations. The resulting protection measure against the THA is entirely passive, thus preventing the loopholes inherent to active, more sophisticated, countermeasures. We believe it will become a standard tool in all quantum-secured optical systems that need to guarantee the protection of a private space.

\section*{Acknowledgments}
\label{sec:7}

\noindent We are indebted to Norbert L\"{u}tkenhaus, for his scientific support throughout the work preparation, and to Kiyoshi Tamaki, for noticing an ill-posed generalization of Ref.~\cite{TCK+14} during his critical reading of the manuscript. Useful discussions with Bernd Fr\"{o}hlich are gratefully acknowledged.

\bigskip

\section*{Appendix}
\label{sec:8}

\subsection{Security-related assumptions}
\label{subsec:8Z}

\noindent In the main text, we considered the unidirectional, fiber-based, phase-modulated QKD setup depicted in Fig.~\ref{fig:2} and studied its resistance to the THA. For that, we have made a number of assumptions that we summarize in the following list:
\begin{enumerate}
  \item Alice has the ability to bound $N$, the number of Trojan photons entering her setup. She can characterize the components in her setup and test whether they behave as expected under all relevant conditions.
  \item Eve uses a tensor product of coherent states to execute the THA. The intensity of the coherent states has not to be constant, but it is advantageous for Eve to choose it so.
  \item Alice's light source emits either single-photon states or phase-randomized coherent states that are perfectly encoded into the states of the BB84 protocol. Imperfect encoding of the initial states as studied, e.g., in Ref.~\cite{TCK+14} is excluded. The only side-channel in the QKD setup is the THA against Alice's phase modulator described in the main body of this work.
  \item The detection efficiency of the receiver is independent of the basis choice and the basis is randomly chosen by the users.
  \item The key rate is worked out in the asymptotic scenario, assuming that Alice and Bob have infinitely many signals and decoy states to generate the key.
  \item The reflectivity measured via the OTDR experiment is a linear function of the input polarization.
\end{enumerate}
Without Assumption~1, it would be impossible to prove security. If the quantity $N$ cannot be bounded, there is no private space for the encoding of the classical information onto the quantum systems, and the quantum protection is circumvented. In the main text, $N$ was bounded using the LIDT of the OFL in Fig.~\ref{fig:6}. It is natural to ask whether a power monitor or a watchdog detector, placed at the entrance of Alice's unit to actively monitor the input power, can provide an alternative, better, bound to $N$. There are reasons suggesting against this option. Firstly, an additional detector would add extra cost and complexity to the setup, opening up additional potential loopholes. For example, it has been shown in~\cite{SRK+14} that a power monitor can be easily bypassed if not properly engineered. Secondly, we used Eq.~\eqref{eq:13} for the LIDT, according to which narrow pulses of light create a larger damage to the optical component than continuous-wave light, so they are more easily detectable by the users. This let us draw a worst-case scenario for Eve's laser. We are not aware of a similar law applicable to a power monitor. Finally, even if the power monitor solution worked fine and allowed to reduce the input photon number $N$ by several orders of magnitude, it should still be compared to the six or more orders of magnitude guaranteed by the addition of a single inexpensive and nearly loss-free component like an optical isolator.

As for the second part of Assumption~1, if Alice cannot characterize her components, she cannot work out the value of the optical isolation $\gamma$ to relate $N$ and $\mu_{\textrm{out}}$ via Eq.~\eqref{eq:3}. The characterization should consider the physical limits imposed to Eve's laser. For example, Eve's laser's power is constrained by the LIDT of the OFL in Fig.~\ref{fig:6}. Therefore the behaviour of the components should be tested up to the LIDT value of the OFL. This excludes hacking strategies leveraging on an unexpected behavior of the real components, passively or actively triggered by the eavesdropper. The characterization step could be simplified if an optical fuse with a LIDT value lower than the lowest tolerance threshold of the components in Alice's setup were available~\cite{NOTE8}.

Assumption~2 allows us to write the states leaving Alice's apparatus as in Eq.~\eqref{eq:7}. It is possible, in principle, that phase-sensitive states of light, e.g. squeezed states~\cite{L73}, could provide Eve with more information than coherent states. However, as Table~I shows, the value of the attenuation in Alice's setup is at least 170~dB. It seems unlikely that the fragile squeezed state can survive in this lossy environment. The second part of this assumption descends from the convexity of the secure key rate as a function of $\mu_{\textrm{out}}$, which has been verified for all the key rates presented in this work.

Assumption~3 is necessary to remove additional side-channels that could, in principle, enhance the THA, e.g., encoding states that are different from the ideal ones prescribed by the BB84 protocol. Also, it guarantees that the rate equations derived for the decoy-state BB84 protocol hold, because Eve's tampering with Alice's decoy state estimation would represent an additional side channel and would contradict the assumption. Extending the security argument to decoy states without making use of Assumption~3 could be a trivial task and a detailed separate study is required.
However, we would like to speculate on this point further.

For simplicity, we assume that the light emitted by Alice is phase-randomized. In some cases, this is simple to guarantee, e.g., when phase-randomization is an intrinsic feature of the light source~\cite{YLD+14}. In other cases, when phase randomization is committed to a separate active component~\cite{ZQL07}, it could be more difficult to show that Eve cannot access this extra component with a more refined THA.
With phase randomization on hand, the decoy state technique requires that the intensity of the emitted light is varied in a random way, known to Alice. This can be achieved by adding an intensity modulator (IM) to the setup of Fig.~\ref{fig:2}, between the interferometer and the laser source. If there is no additional optical isolation between the IM and the interferometer, the optical isolation $\gamma$ that shields Alice's PM from Eve applies to the IM too and the coherent state sent by Eve to probe the IM returns to her with an average photon number not larger than $\mu_{\textrm{out}}$. 
However, if there is a perfect optical isolator between the IM and the interferometer, then the Trojan photons retrieved by Eve are only informative about Alice's PM, whereas the IM is perfectly shielded from Eve. This latter case is an example of how the Assumption~3 can be enforced.
However, because a perfect isolation is impossible in practice, we considered how the key rates of the decoy-state BB84 would change if a \textit{single real} optical isolator, guaranteeing 50~dB isolation, were used instead. In this case, the $\mu_{\textrm{out}}$ back-reflected to Eve from the IM would be 5 orders of magnitude smaller than the one back-reflected from the PM. Applying to this realistic scenario an argument similar to the one described in the forthcoming Appendix~C.2, we found key rates that are indistinguishable from the ones presented in this work.

Assumption 4 and 5 are related to the proof methods adopted by us to draw the key rates in presence of a THA~\cite{GLLP04,K09,LP07,TCK+14}. There, security was proven in the asymptotic scenario leveraging on the fact that the measurement performed by the receiver is equivalent to a basis-independent filter followed by a two-valued POVM (Positive-Operator Valued Measure). In~\cite{K06} it was shown that this assumption can be enforced if Bob's single-photon detectors have equal efficiency and if their dark counts and efficiencies are carefully modeled. The detectors can be threshold detectors and in this case a specific value of the key bit must be assigned whenever both detectors click, to guarantee the basis-independence condition.

Assumption 6 is necessary during the characterization stage to upper bound the reflectivity of the transmitter, as shown in Sec.~\ref{sec:4}.A. To meet this assumption, we put particular care in the OTDR experiment to avoid nonlinear effects~\cite{Boy08} due to a high power from the laser, which is the only source of light in the experiment. For that, the intensity of the laser was set to about 6~nW. Let us notice that Eve is not playing any role here, because the characterization of the QKD setup is accomplished in a protected environment. Therefore, we can safely assume that the reflectivity depends linearly on the polarization, as in ordinary Fresnel equations.

\subsection{Rate equations for the Trojan-horse attack: general case}
\label{subsec:8A}

\noindent With the assumptions of the previous section on hand, let us describe the security argument in more detail.
In the GLLP-Koashi approach~\cite{GLLP04,K09}, an entanglement-based description of the preparation stage is adopted.
The states to be prepared are given in Eq.~\eqref{eq:7}. We rewrite them here for convenience:
\begin{eqnarray}
\label{eq:7bis}
\nonumber \ket{\psi_{0X}}_{BE} &=& \ket{0_X}_B \otimes \ket{+\sqrt{\mu_{\textrm{out}}}}_E  , \\
\nonumber \ket{\psi_{1X}}_{BE} &=& \ket{1_X}_B \otimes \ket{-\sqrt{\mu_{\textrm{out}}}}_E  , \\
\nonumber \ket{\psi_{0Y}}_{BE} &=& \ket{1_Y}_B \otimes \ket{+i\sqrt{\mu_{\textrm{out}}}}_E , \\
          \ket{\psi_{1Y}}_{BE} &=& \ket{0_Y}_B \otimes \ket{-i\sqrt{\mu_{\textrm{out}}}}_E .
\end{eqnarray}
The $X$ basis states of Eq.~\eqref{eq:7bis} can be prepared by Alice by measuring in the basis $\{|0_X\rangle_A , |1_X\rangle_A \}$ the following entangled state:
\begin{equation}
\label{eq:19}
    \ket{\Psi_X} = \frac{\ket{0_X}_A \ket{\psi_{0X}}_{BE} + \ket{1_X}_A \ket{\psi_{1X}}_{BE} }{\sqrt{2}}.
\end{equation}
Similarly, the $Y$ basis states of Eq.~\eqref{eq:7bis} can be prepared by measuring in the basis $\{|0_Y\rangle_A , |1_Y\rangle_A \}$ the state:
\begin{equation}
\label{eq:20}
    \ket{\Psi_Y} = \frac{\ket{0_Y}_A \ket{\psi_{0Y}}_{BE} + \ket{1_Y}_A \ket{\psi_{1Y}}_{BE} }{\sqrt{2}}.
\end{equation}
If the state preparation stage was perfect, the two states $\ket{\Psi_X}$ and $\ket{\Psi_Y}$ would be indistinguishable, as it can be verified from the above equations in the limit $\mu_{\textrm{out}}\rightarrow 0$. In this case, we know that the secure key rate of the single-photon efficient BB84 protocol with data basis $X$ and test basis $Y$ would be:
\begin{equation}
\label{eq:RSP}
  R_{\textrm{ideal}} = Q_X [1-h(e_Y)-f_{EC} h(e_X)] ,
\end{equation}
where $Q_X$ is the single-photon detection rate in the $X$ basis; $e_Y$ ($e_X$) is the error rate measured from single photons in the $Y$ ($X$) basis; $f_{EC} \geq 1$ is the inefficiency of error correction~\cite{NOTE5}. Because we are considering here a single-photon source, all the quantities in Eq.~\eqref{eq:RSP} refer to the single photon case.

When the preparation is not perfect, or part of the basis information leaks out of the transmitting unit, the states $\ket{\Psi_X}$ and $\ket{\Psi_Y}$ are different and the above key rate has to be replaced by the following one~\cite{LP07}:
\begin{equation}
\label{eq:RSP2}
  R = Q_X [1-h(e'_Y)-f_{EC} h(e_X)].
\end{equation}
In Eq.~\eqref{eq:RSP2}, the phase error rate $e_Y$ has been replaced by a larger error rate, $e'_Y \geq e_Y$. It was shown in~\cite{K09} that the term $e'_Y$ is an upper bound to the error rate that the users would find if they measured the $X$-basis state $\ket{\Psi_{X}}$ in the basis $Y$.

To find the relation between the error rates $e'_Y$ and $e_Y$, we can imagine that Alice owns a private bidimensional quantum system, a ``quantum coin''~\cite{GLLP04}, and prepares the following state:
\begin{equation}
\label{eq:21}
    \ket{\Phi} = \frac{\ket{0_Z}_{C} \ket{\Psi_{X}} + \ket{1_Z}_{C} \ket{\Psi_{Y}} }{\sqrt{2}},
\end{equation}
where the subscript $C$ refers to the quantum coin.
The states in Eqs.~\eqref{eq:7bis} can then be prepared by Alice by first measuring the quantum coin in the basis $\{\ket{0_Z}_C, \ket{1_Z}_C\}$ and then, depending on the outcome, measuring the resulting state $\ket{\Psi_{X}}$ or $\ket{\Psi_{Y}}$ in the basis $\{\ket{0_X}_A, \ket{1_X}_A\}$ or $\{\ket{0_Y}_A, \ket{1_Y}_A\}$, respectively. Because Eve has no access to the quantum coin, she cannot distinguish this virtual preparation from the real preparation executed in the actual protocol. Therefore we are allowed to think that Alice prepares her initial states using the quantum coin.
Also, she can delay her measurement until after Bob has measured the states received from Alice. In this scenario, by noting that Eve's information about Alice's key does not change if Bob measures $\ket{\Psi_{X}}$ in the basis $Y$, Koashi obtained $e'_Y$ from $e_Y$ using a complementarity argument, by applying the ``Bloch sphere bound''~\cite{TKI03} to the quantum coin~\cite{K09}.

Let us quantify the basis dependence of Alice's states in terms of the quantum coin imbalance~\cite{GLLP04}. By rewriting Eq.~\eqref{eq:21} in the $X$ basis of the quantum coin, we find:
\begin{equation}
\label{eq:21bis}
    \ket{\Phi} = \frac{ \ket{0_X}_{C} ( \ket{\Psi_{X}} + \ket{\Psi_{Y}} ) + \ket{1_X}_{C} ( \ket{\Psi_{X}} - \ket{\Psi_{Y}} ) }{2}.
\end{equation}
To quantify the basis dependence of Alice's states, we need to evaluate the probability that the two states $\ket{\Psi_{X}}$ and $\ket{\Psi_{Y}}$ are different. From the above equation, it amounts to the probability that Alice obtains the outcome $X=-1$, associated to the state $\ket{1_X}_{C}$, when she measures the quantum coin in the basis $X$. We call $\Delta$ this probability:
\begin{equation}
\label{eq:Delta}
    \Delta = \textrm{Prob}(X_C=-1)=\frac{1- \textrm{Re}( \langle \Psi_{X} | \Psi_{Y} \rangle)}{2}.
\end{equation}
Let us estimate it for the states prepared by Alice. From Eqs.~\eqref{eq:19} and \eqref{eq:20} we can calculate:
\beq \label{eq:23}
\Delta = \frac{1}{2} [1-\exp(-\mu_{\textrm{out}}) \cos(\mu_{\textrm{out}})].
\eeq
When $\mu_{\textrm{out}}=0$, $\Delta=0$ and the states emitted by Alice are basis independent. However, when $\mu_{\textrm{out}}>0$, $\Delta$ is positive and the states carry some basis information out of Alice's enclosure. The basis information can be exploited by Eve to enhance her strategy, acting on the channel losses, which are entirely under her control. Specifically, she can replace the real channel with another, loss-free, channel. Then she selectively stops all the states that are not favorable to her, until the loss rate measured by the users is matched. To account for this possibility, the users must consider the worst case where all the non-detected events were coming from $X=1$ eigenstates of the quantum coin and renormalize $\Delta$ accordingly:
\begin{equation}
\label{eq:24}
    \Delta' = \frac{\Delta}{\mathcal{Y}}.
\end{equation}
In Eq.~\eqref{eq:24}, $\mathcal{Y}=\min(\mathcal{Y}_X,\mathcal{Y}_Y)$, with $\mathcal{Y}_X$  and $\mathcal{Y}_Y$ the single-photon yields measured in the $X$  and $Y$ basis, respectively.
Finally, using the Bloch sphere bound~\cite{TKI03} and the effective coin imbalance $\Delta'$, the relation between the phase error rates $e'_Y$ and $e_Y$ is obtained as~\cite{K09,LP07}:
\bea \label{eq:25bis}
\nonumber   e'_Y &=& e_Y + 4 \Delta' (1-\Delta') (1-2e_Y)+ \\
            &+& 4(1-2\Delta')\sqrt{\Delta' (1-\Delta') e_Y (1-e_Y)}.
\eea
When the single-photon source is replaced by a decoy-state source and under Assumption~3 of Appendix~A, the resulting rate is a straightforward generalization of Eq.~\eqref{eq:RSP2} along the lines described in~\cite{LMC05}. Indicating with a tilde the quantities to be estimated via the decoy-state technique, we have:
\begin{equation}
\label{eq:RSP3app}
  \widetilde{R} = \widetilde{Q}_X^{(1)}  \left\{ 1-h \left[ \tilde{e}^{\prime(1)}_Y \right] \right\} - Q_X^{(s)} f_{EC} h[e_X^{(s)}],
\end{equation}
where:
\bea \label{eq:25quatris}
\nonumber   \tilde{e}_Y^{\prime (1)} &=& \tilde{e}_Y + 4 \widetilde{\Delta}' (1-\widetilde{\Delta}') (1-2 \tilde{e}_Y)+ \\
\nonumber                &+& 4(1-2 \widetilde{\Delta}')\sqrt{\widetilde{\Delta}' (1-\widetilde{\Delta}') \tilde{e}_Y (1-\tilde{e}_Y)},\\
            \widetilde{\Delta}' &=& \frac{\Delta}{\widetilde{\mathcal{Y}}}.
\eea
In Eq.~\eqref{eq:RSP3app}, $\widetilde{Q}_X^{(1)}$ is the decoy-state estimation of the single-photon detection rate $Q_X$ (see Eq.~\eqref{eq:RSP2}) and $Q_X^{(s)}$ is the detection rate of the signal pulse measured in the $X$ basis.
In Eq.~\eqref{eq:25quatris}, we conservatively defined $\widetilde{\mathcal{Y}}=\min[\widetilde{\mathcal{Y}}_X,\widetilde{\mathcal{Y}}_Y]$, with $\widetilde{\mathcal{Y}}_X$  and $\widetilde{\mathcal{Y}}_Y$ the single-photon yields in the $X$ and $Y$ basis, respectively, estimated via the decoy state technique.

\subsection{Rate equations for two specific Trojan-horse attacks}
\label{subsubsec:8B}

\subsubsection{Trojan-horse attack with a passive use of the Trojan photons}
\label{subsubsec:8Ba}

\noindent We analyze the security of the BB84 protocol against a different, less general, THA. This serves a twofold purpose: it provides un upper bound to the key rate achievable in presence of a THA and gives us a chance to use a different proof method to study the THA.

In the specific THA of this section, Eve uses the information leaked from Alice in a passive way. She extracts from the quantum channel the states labelled with $E$ in Eq.~\eqref{eq:7bis} and stores them in a perfect quantum memory. This causes no disturbance on the quantum channel connecting Alice and Bob. Then, during the basis reconciliation stage of the BB84 protocol, Eve learns the basis information communicated by the users on a public channel. This allows her to measure the stored states in the same bases as the users and learn the resulting key bit every time the result of her measurement is conclusive. The conclusiveness of her results depends on the magnitude of the parameter $\mu_{\textrm{out}}$ in the stored states.

We analyze this situation using the \textit{loss-tolerant} proof method described by Tamaki~\textit{\underline{et al}} in~\cite{TCK+14}. For that, we can use the equations of the previous section until Eq.~\eqref{eq:RSP2}. The difference starts with the estimation of the phase error rate in the virtual protocol, $e'_Y$, which is more direct than in the GLLP-Koashi approach.

We consider a \textit{real} protocol, in which Alice prepares the state in Eq.~\eqref{eq:7bis} and sends them to Bob (and Eve), and a \textit{virtual} protocol, in which an entanglement-based view is adopted. In both cases, we assume that Bob's measurement does not depend on the basis choice. This is guaranteed by the Assumption 4 discussed in Appendix~A.

In the virtual protocol, Alice prepares the states to be sent to Bob by measuring her half of an entangled state. This is the same state as in Eq.~\eqref{eq:19}, which we rewrite here both in the $X$ and in the $Y$ basis:
\begin{equation}
\label{eq:27}
    \ket{\Psi} = \frac{\ket{0_X}_A \ket{\psi_{0X}}_{BE} + \ket{1_X}_A \ket{\psi_{1X}}_{BE} }{\sqrt{2}} ,
\end{equation}
\begin{equation}
\label{eq:28}
    \ket{\Psi} = \frac{\ket{0_Y}_A \ket{\phi_{1Y}}_{BE} + \ket{1_Y}_A \ket{\phi_{0Y}}_{BE} }{\sqrt{2}} .
\end{equation}
In Eqs.~\eqref{eq:27} and \eqref{eq:28} we have defined:
\begin{eqnarray}
\label{eq:29}   \ket{\phi_{1Y}}_{BE} &=& \frac{-i\ket{0_Y}_B \ket{\epsilon_{-}}_{E} + \ket{1_Y}_B \ket{\epsilon_{+}}_{E} }{\sqrt{2}} , \\
\label{eq:29b}  \ket{\phi_{0Y}}_{BE} &=& \frac{\ket{0_Y}_B \ket{\epsilon_{+}}_{E} + i \ket{1_Y}_B \ket{\epsilon_{-}}_{E} }{\sqrt{2}} , \\
\label{eq:29eps}\ket{\epsilon_{\pm}} &=& \frac{\ket{\sqrt{\mu_{\textrm{out}}}} \pm \ket{-\sqrt{\mu_{\textrm{out}}}} }{\sqrt{2}} .
\end{eqnarray}
Notice that when $\mu_{\textrm{out}} \rightarrow 0$, $\ket{\psi_{0X}}_{BE}\rightarrow \ket{0_X}_B\ket{v}_E$, $\ket{\psi_{1X}}_{BE}\rightarrow \ket{1_X}_B\ket{v}_E$, $\ket{\phi_{1Y}}_{BE}\rightarrow \ket{1_Y}_B\ket{v}_E$ and $\ket{\phi_{0Y}}_{BE}\rightarrow \ket{0_Y}_B\ket{v}_E$, where $\ket{v}$ is the vacuum state, thus recovering from Eqs.~\eqref{eq:27} and \eqref{eq:28} two maximally entangled states in a two-dimensional Hilbert space tensor product with the vacuum state. This situation is secure against the THA and constitutes a reference for our later argument in Appendix~C.2. However, when $\mu_{\textrm{out}} > 0$, the effective Hilbert space's dimension becomes larger than two, favoring the THA. %
Note also that the states in Eq.~\eqref{eq:29eps} are orthogonal but not normalized, while the states in Eqs.~\eqref{eq:29} and~\eqref{eq:29b} are normalized. More specifically: $\braket{\epsilon_{\pm}}{\epsilon_{\pm}}=1 \pm \exp(-2 \mu_{\textrm{out}})$, $\braket{\epsilon_{\pm}}{\epsilon_{\mp}}=0$, $\braket{\phi_{wY}}{\phi_{wY}}=1$, with $w=\{0,1\}$.

Let us assume for the moment that the system $E$ is not accessible either to Alice or to Eve. Under this assumption, the proof method in Ref.~\cite{TCK+14} applies. This is so for a twofold reason. First, the security argument in~\cite{TCK+14} is based on the description given in~\cite{K09,LP07}, which allows for an enlarged dimension of Alice's Hilbert space. Second, Eve cannot perform a basis-dependent selection of the states emitted by Alice, because the basis information is contained in the system $E$, which is not accessible to her. Therefore, as shown in~\cite{TCK+14}, she cannot modify the transmission rates of Alice's states using the basis information potentially leaked from Alice's module.
We notice that also in the specific THA considered here Eve has no chance to modify the transmission rates during the quantum transmission, due to the fact that Eve is allowed to access the system $E$ only \textit{after} the basis information has been publicly disclosed by the users.

Given that, suppose that Alice measures the ancillary states of $\ket{\Psi}$ in the $Y$ basis. Because the states in Eqs.~\eqref{eq:29} and~\eqref{eq:29b} are normalized, she will obtain with probability $1/2$ the state $\ket{0_Y}$ and with probability $1/2$ the state $\ket{1_Y}$, thus projecting $\ket{\Psi}$ into one of the following two states, respectively:
\begin{eqnarray}
\label{eq:30}
\nonumber   \rho_B^{(0)}&=& \textrm{Tr}_E (\ket{\phi_{1Y}}_{BE} \bra{\phi_{1Y}}) \\
\nonumber               &=& c_{-} \ket{0_Y}_B \bra{0_Y} + c_{+} \ket{1_Y}_B \bra{1_Y} \\
                        &=& \frac{1 }{2} [ \hat{\sigma}_0 - \exp(-2 \mu_{\textrm{out}}) \hat{\sigma}_2 ] ,
\end{eqnarray}
\begin{eqnarray}
\label{eq:31}
\nonumber   \rho_B^{(1)}&=& \textrm{Tr}_E (\ket{\phi_{0Y}}_{BE} \bra{\phi_{0Y}}) \\
\nonumber               &=& c_{+} \ket{0_Y}_B \bra{0_Y} + c_{-} \ket{1_Y}_B \bra{1_Y} \\
                        &=& \frac{1}{2} [ \hat{\sigma}_0 + \exp(-2 \mu_{\textrm{out}}) \hat{\sigma}_2 ]  ,
\end{eqnarray}
where we have defined $c_{\pm}:=\braket{\epsilon_{\pm}}{\epsilon_{\pm}} /2$ and introduced the identity operator in the $2$-dimensional Hilbert space $\hat{\sigma}_0$ and the Pauli matrix $\hat{\sigma}_2=[(0,-i),(i,0)]$. These operators are necessary to connect the $Y$-basis states of the virtual protocol, Eq.~\eqref{eq:29} and \eqref{eq:29b}, to the $Y$-basis states of the real protocol, contained in the third and fourth line of Eq.~\eqref{eq:7bis}. Because any qubit state can be written as a linear combination of identity and Pauli matrices, its transmission rate can be obtained directly from the Pauli matrices' transmission rates~\cite{TCK+14}.
Accordingly, we define the transmission rate of $\hat{\sigma}_k$, $k=\{0,2\}$, as $q_{s_Y|k}=\textrm{Tr}(\hat{D}_{s_Y} \hat{\sigma}_k )/2$, with $\hat{D}_{s_Y}=\sum_l \hat{A}^\dag_l \hat{M}_{s_Y} \hat{A}_l$, $\hat{A}_l$ an arbitrary operator associated to Eve's action and $\hat{M}_{s_Y}$ the operator representing Bob's POVM in the $Y$ basis associated to the bit value $s$. We can then obtain the transmission rates in the virtual and real protocol as combinations of the $q_{s_Y|k}$'s.

Let us call $p_y$ the probability that Alice and Bob both select the $Y$ basis. In the real protocol (superscript $r$), the joint probability $\mathcal{P}_{j_Y, i_Y}$ that Alice sends out the state $\ket{i_Y}$ and Bob detects $\ket{j_Y}$ ($i,j={0,1}$) is for each pair of states:
\begin{eqnarray}
\label{eq:32}
\nonumber   \mathcal{P}_{0_Y,1_Y}^{(r)} &=& \frac{p_y^2}{2}(q_{0_Y|0} + q_{0_Y|2}) , \\
\nonumber   \mathcal{P}_{1_Y,1_Y}^{(r)} &=& \frac{p_y^2}{2}(q_{1_Y|0} + q_{1_Y|2}) , \\
\nonumber   \mathcal{P}_{0_Y,0_Y}^{(r)} &=& \frac{p_y^2}{2}(q_{0_Y|0} - q_{0_Y|2}) , \\
            \mathcal{P}_{1_Y,0_Y}^{(r)} &=& \frac{p_y^2}{2}(q_{1_Y|0} - q_{1_Y|2}) .
\end{eqnarray}
The corresponding probabilities in the virtual protocol (superscript $v$) are:
\begin{eqnarray}
\label{eq:33}
\nonumber   \mathcal{P}_{0_Y,1_Y}^{(v)} &=& \frac{p_y^2}{2}(q_{0_Y|0} + e^{-2 \mu_{\textrm{out}}} q_{0_Y|2}) , \\
\nonumber   \mathcal{P}_{1_Y,1_Y}^{(v)} &=& \frac{p_y^2}{2}(q_{1_Y|0} + e^{-2 \mu_{\textrm{out}}} q_{1_Y|2}) , \\
\nonumber   \mathcal{P}_{0_Y,0_Y}^{(v)} &=& \frac{p_y^2}{2}(q_{0_Y|0} - e^{-2 \mu_{\textrm{out}}} q_{0_Y|2}) , \\
            \mathcal{P}_{1_Y,0_Y}^{(v)} &=& \frac{p_y^2}{2}(q_{1_Y|0} - e^{-2 \mu_{\textrm{out}}} q_{1_Y|2}) .
\end{eqnarray}
In order to define the phase error rate, we need to identify the error event in the virtual protocol. This can be done using Eqs.~\eqref{eq:7bis}, \eqref{eq:28}, \eqref{eq:29} in the limit $\mu_{\textrm{out}}\rightarrow 0$. When there is no THA on the channel, Bob obtains the correct state $\ket{1_Y}$ ($\ket{0_Y}$) when Alice measures $\ket{0_Y}$ ($\ket{1_Y}$) on her ancillary states. Hence we associate an error with both Alice and Bob obtaining the same state, $\ket{0_Y}$ or $\ket{1_Y}$. So the phase error rate can be written as:
\begin{equation}
\label{eq:34}
    e'_Y = \frac{ \mathcal{P}_{0_Y,0_Y}^{(v)} + \mathcal{P}_{1_Y,1_Y}^{(v)} }{\mathcal{P}_{0_Y,0_Y}^{(v)} + \mathcal{P}_{0_Y,1_Y}^{(v)} + \mathcal{P}_{1_Y,0_Y}^{(v)} + \mathcal{P}_{1_Y,1_Y}^{(v)} } .
\end{equation}
From Eqs.~\eqref{eq:32}, \eqref{eq:33} we can rewrite the phase error rate in terms of the rates measured in the real protocol. The result is:
\beq \label{eq:35}
e'_Y = \frac{1}{2} [1-a_{\mathcal{P}}^{(r)} \exp(-2 \mu_{\textrm{out}})] ,
\eeq
where we have set:
\begin{equation}
\label{eq:36}
    a_{\mathcal{P}}^{(r)} = \frac{\sum_{i,j=\{0,1\}}(-)^{i+j+1} \mathcal{P}_{j_Y,i_Y}^{(r)}}{\sum_{i,j=\{0,1\}} \mathcal{P}_{j_Y,i_Y}^{(r)}}.
\end{equation}
The secure key rate is obtained by replacing the phase error of Eq.~\eqref{eq:35} into Eq.~\eqref{eq:RSP2}:
\begin{equation}
\label{eq:9}
    R^* = Q_X \left[  1-h\left(  e'_Y \right) -f_{EC}h\left(  e_{X}\right)  \right].
\end{equation}
The key rate in Eq.~\eqref{eq:9} applies to slightly more general THA than the specific one considered in this section. It applies to all THA in which Eve cannot interact with the auxiliary Trojan-horse states (labelled with $E$ in Eq.~\eqref{eq:7bis}~) \textit{during} the transmission of the qubit states (labelled with $B$ in Eq.~\eqref{eq:7bis}~).
We already noted that if Eve cannot access the auxiliary system $E$ during the quantum transmission, she cannot selectively modify the transmission rates $\mathcal{P}$. Here we additionally note that even if Eve changed her action, described by the operators $\hat{A}_l$, according to whether she will own or not the auxiliary system $E$ after the basis reconciliation step, she would not gain more information about the final key. This descends from Koashi's proof method~\cite{K09}, upon which the proof described in~\cite{TCK+14} is built. There, it was shown that irrespective of who owns the auxiliary system, whether Alice or Eve, if the users can obtain a faithful estimation of the phase error rate $e'_Y$, they can in principle distil a perfect qubit in a $Y$ eigenstate. When measured by Alice in the data basis $X$, the $Y$ eigenstate always provides her with a fully random key bit, not predictable by Eve. Therefore, even if Eve tunes her choice of the operators $\hat{A}_l$ on the auxiliary system $E$, her knowledge of the final key does not increase. The only condition required is that Eve accesses the auxiliary system $E$ \textit{after} the quantum transmission has been completed by the users.
\begin{figure}[tbp]
%
  \includegraphics[width=8.0cm]{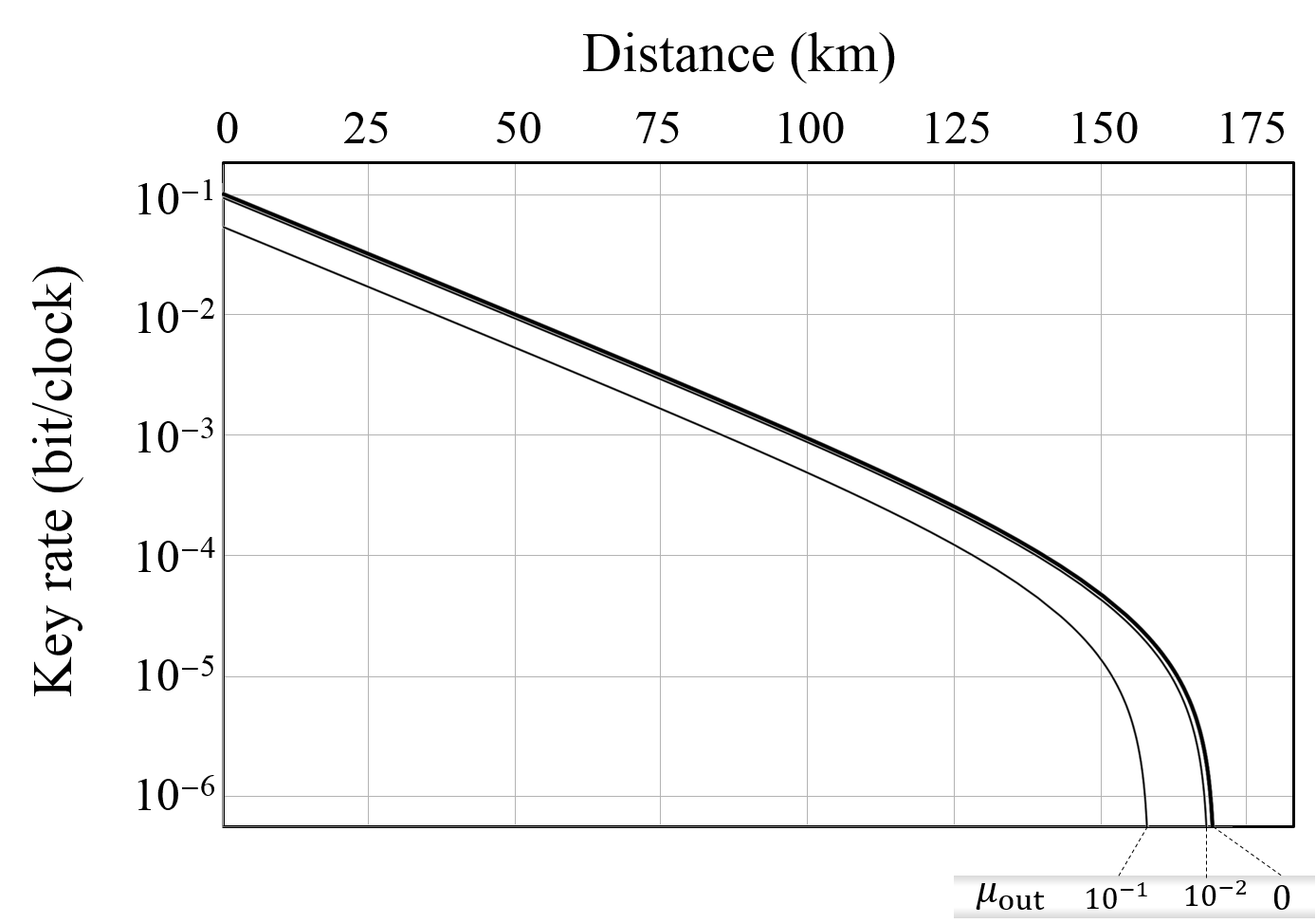}\\
  \caption{Asymptotic key rate $R^*$ versus distance for the single-photon efficient BB84 protocol, under a passive THA. The rate is plotted for various values of the parameter $\mu_{\textrm{out}}$. Parameters in the simulation are as in Fig.~\ref{fig:3}. }
  \label{fig:4bis}
\end{figure}
\begin{figure}[tbp]
%
  \includegraphics[width=8.0cm]{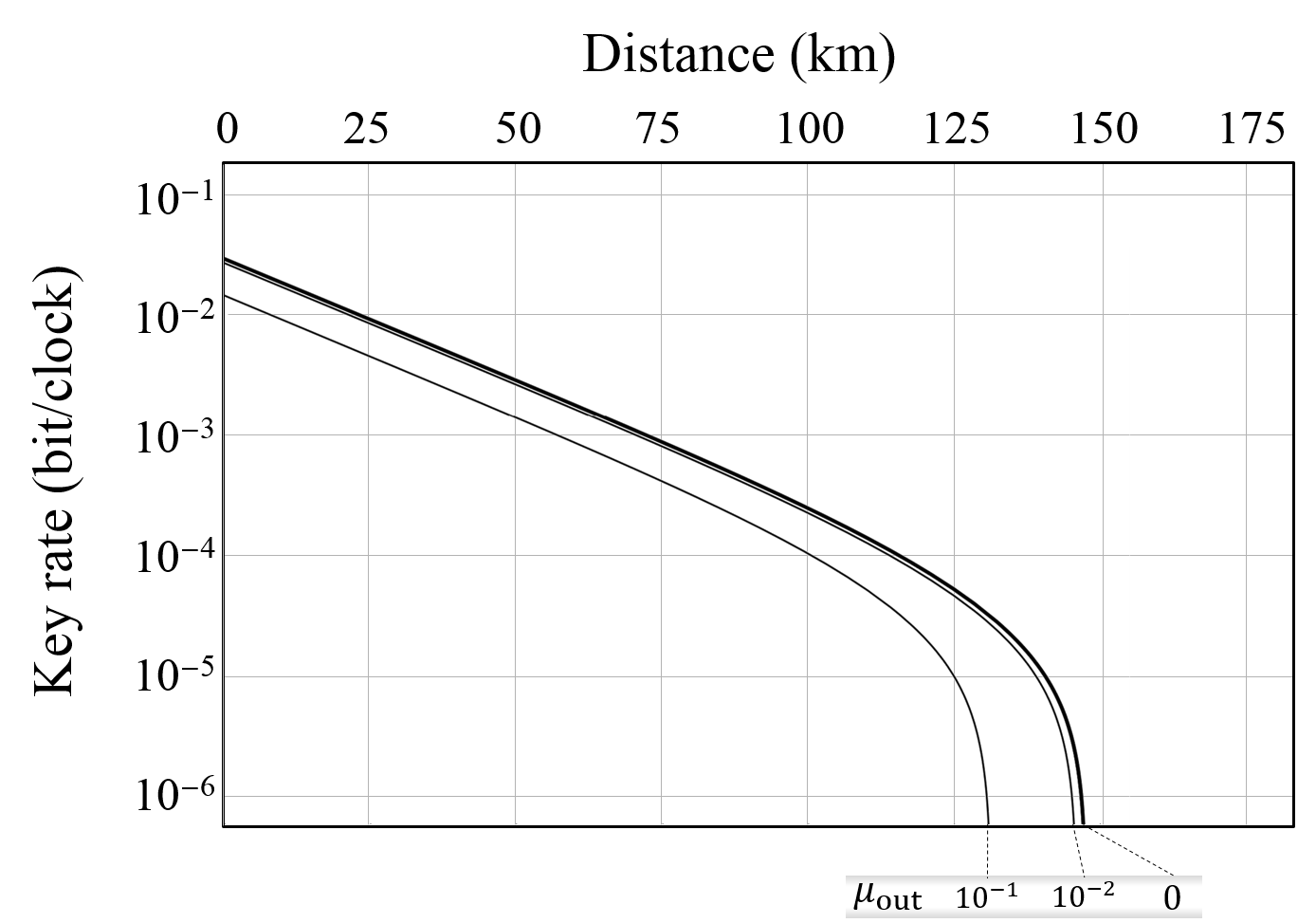}\\
  \caption{Asymptotic key rate $\widetilde{R}^*$ versus distance for the decoy-state efficient BB84 protocol, under a passive THA. The rate is plotted for various values of the parameter $\mu_{\textrm{out}}$. Parameters in the simulation are as in Fig.~\ref{fig:4}. }
  \label{fig:4tris}
\end{figure}

To adapt the key rate in Eq.~\eqref{eq:9} to the decoy-state estimation technique we exploit Assumption~3 in Appendix~A, according to which Eve cannot use the THA to modify the decoy-state estimation. Then we need to show which quantities have to be estimated using decoy states. We use the tilde to explicitly indicate such quantities in the key rate:
\beq \label{eq:12bis}
      \widetilde{R}^* = \widetilde{Q}_X^{(1)}  \left\{ 1-h \left[\tilde{e}^{\prime(1)}_Y \right] \right\} - Q_X^{(s)} f_{EC} h[e_X^{(s)}] ,
\eeq
where $s$ is the mean photon number of the signal in the decoy-state set, $\widetilde{Q}_X^{(1)}$ is the overall single-photon detection rate in the $X$-basis, estimated using the decoy-state technique, $Q_X^{(s)}$ and $e_X^{(s)}$ are, respectively, the measured detection and error rates for the signal in the $X$ basis. Furthermore, we have set:
\bea \label{eq:12bisbis}
\nonumber   \tilde{e}^{\prime(1)}_Y &=& \frac{1}{2} [1 - a_{\widetilde{\mathcal{P}}}^{(r)} \exp (-2 \mu_{\textrm{out}}) ]~, \\
            a_{\widetilde{\mathcal{P}}}^{(r)} &=& \frac{\sum_{i,j=\{0,1\}}(-)^{i+j+1} \widetilde{\mathcal{P}}_{j_Y,i_Y}^{(r)}}{\sum_{i,j=\{0,1\}} \widetilde{\mathcal{P}}_{j_Y,i_Y}^{(r)}},
\eea
which are a straightforward generalization of Eqs.~\eqref{eq:35} and~\eqref{eq:36}.

The key rates $R^*$ and $\widetilde{R}^*$ are plotted in Figs.~\ref{fig:4bis} and~\ref{fig:4tris}. Both the resulting key rates show no strong dependence on the mean Trojan photon number $\mu_{\textrm{out}}$. The key rates are coincident with the ideal rate corresponding to no THA for all values of $\mu_{\textrm{out}}$ smaller than $\sim 10^{-2}$, and remains positive up to values $0.5$ ($0.38$) in case of a single-photon (decoy-state) source. This represents an improvement of several orders of magnitude over the key rates for a general THA presented in Sec.~\ref{sec:2} and suggests that the power of a THA comes from Eve's capability of selectively introducing losses in the transmission channel, conditional on the information gained from the THA. This observation motivates the study of a more involved THA, in which the shield system $E$ is actively used.

\subsubsection{Trojan-horse attack with active unambiguous state discrimination of the Trojan photons}
\label{subsubsec:8Bb}

\noindent We consider a particular THA in which Eve can access the ancillary system $E$, generated by the THA, during the quantum transmission, i.e., \textit{before} the bases are revealed by the users. However, she can only measure it using a specific measurement described later on. This is more powerful than the THA considered in the previous section, but less powerful than the most general THA discussed in Sec.~\ref{sec:2} and Appendix~B.

In this THA, Eve accesses the space $E$ of the Trojan photons during the quantum transmission stage. She then uses Unambiguous State Discrimination (USD)~\cite{Iva87} to distinguish $\ket{+\sqrt{\mu_{\textrm{out}}}}_E$ from $\ket{-\sqrt{\mu_{\textrm{out}}}}_E$. These states correspond to Alice's states in the data basis ($X$), as per Eq.~\eqref{eq:7bis}. Therefore, whenever the USD succeeds, Eve knows the key bit without measuring, and hence perturbing, Alice's qubit.

However the USD measurement not always provides Eve with a conclusive result and Eve's strategy can be improved as follows. When the result is conclusive, Eve transmits Alice's pulse to Bob without modification; when it is inconclusive, Eve stops Alice's pulse and introduces a loss in the communication channel. Later on, during the basis reconciliation step, Eve will learn the bases chosen by Alice and Bob. After discarding the outcomes of the USD performed on Alice's $Y$-basis states, Eve will be left, ideally, with the same key bits as the users, distilled from the $X$ basis, without having caused any noise on the communication channel.

Let us add more details to this scenario.
When Alice prepares a $Y$-basis state, Eve's retrieved Trojan pulse is in a state $\ket{\pm i \sqrt{\mu_{\textrm{out}}}}_E$. This state cannot help Eve deciding between the two outcomes related to the $X$ basis, because it is equally likely to be projected on either of the two $X$ basis states $\ket{\pm \sqrt{\mu_{\textrm{out}}}}_E$. Therefore, Eve's decision to retain or transmit Alice's state is not related to an increased information gained by Eve and does not require an increase of the privacy amplification performed by the users.
On the contrary, when Alice prepares a $X$-basis state, Eve can modify the transmission rates in a way that affects the security of the system. In a worst-case scenario, we then assume that all the counts detected by Bob come from the $X$ basis and from a conclusive outcome of Eve's USD measurement.

Let us call $p_{\textrm{con}}$ and $p_{\textrm{inc}}=1-p_{\textrm{con}}$ the probabilities of a conclusive and inconclusive outcome, respectively, from the USD of $X$-basis states.
A lower bound to $p_{\textrm{inc}}$ is given by the Ivanovic-Dieks-Peres bound~\cite{Iva87,Die88,Per88}:
\bea
\nonumber p_{\textrm{inc}} &\geq& |\langle \sqrt{\mu_{\textrm{out}}}  |  -\sqrt{\mu_{\textrm{out}}} \rangle |\\
                            &=& \exp(-2 \mu_{\textrm{out}}).
\eea
Then, according to the above-described THA, the fraction of detected events in the $X$ basis that have been transmitted conditional on a conclusive result by Eve is at most:
\beq \label{eq:del}
 \delta \leq \frac{1-p_{\textrm{inc}}}{\mathcal{Y}_X} \leq \frac{1-\exp(-2 \mu_{\textrm{out}})}{\mathcal{Y}},
\eeq
with $\mathcal{Y}:=\min[\mathcal{Y}_X,\mathcal{Y}_Y]$ and $\mathcal{Y}_X$ ($\mathcal{Y}_Y$) the single-photon yield in the $X$ ($Y$) basis.
The fraction $\delta$ (respectively $1-\delta$) contains insecure (secure) bits distilled by the users, because they come from Eve's conclusive (inconclusive) measurement. When the USD is inconclusive, Eve cannot modify selectively Alice's pulses using the system $E$. Therefore, following the same reasoning as in the previous section, we can apply the proof of Ref.~\cite{TCK+14} to this fraction of pulses to estimate the key rate.
\begin{figure}[tbp]
%
  \includegraphics[width=8.0cm]{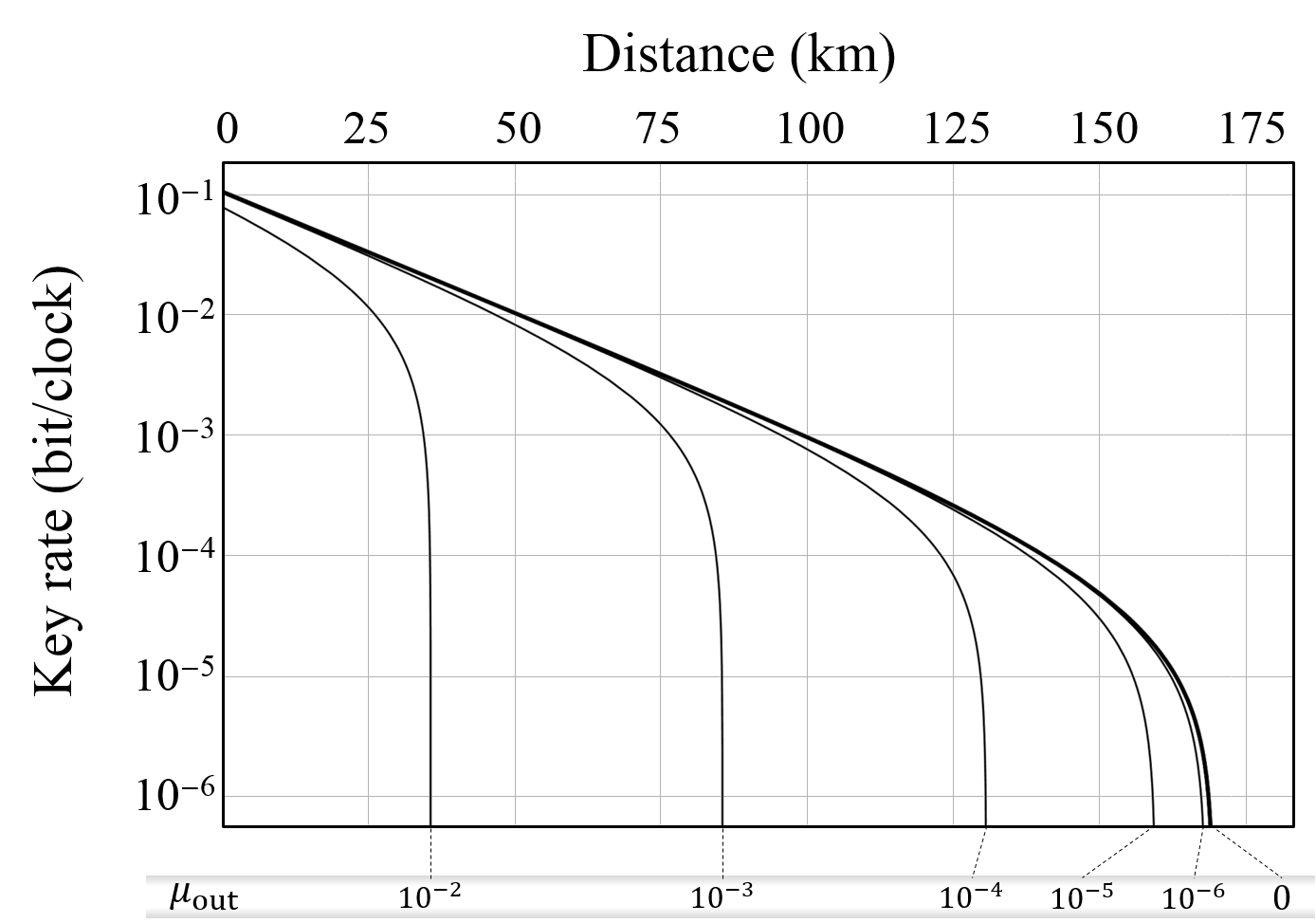}\\
  \caption{Asymptotic key rate $R^{**}$ versus distance for the single-photon efficient BB84 protocol, under a THA with unambiguous state discrimination by Eve. The rate is plotted for various values of the parameter $\mu_{\textrm{out}}$. Parameters in the simulation are as in Fig.~\ref{fig:3}. }
  \label{fig:10a}
\end{figure}
\begin{figure}[tbp]
%
  \includegraphics[width=8.0cm]{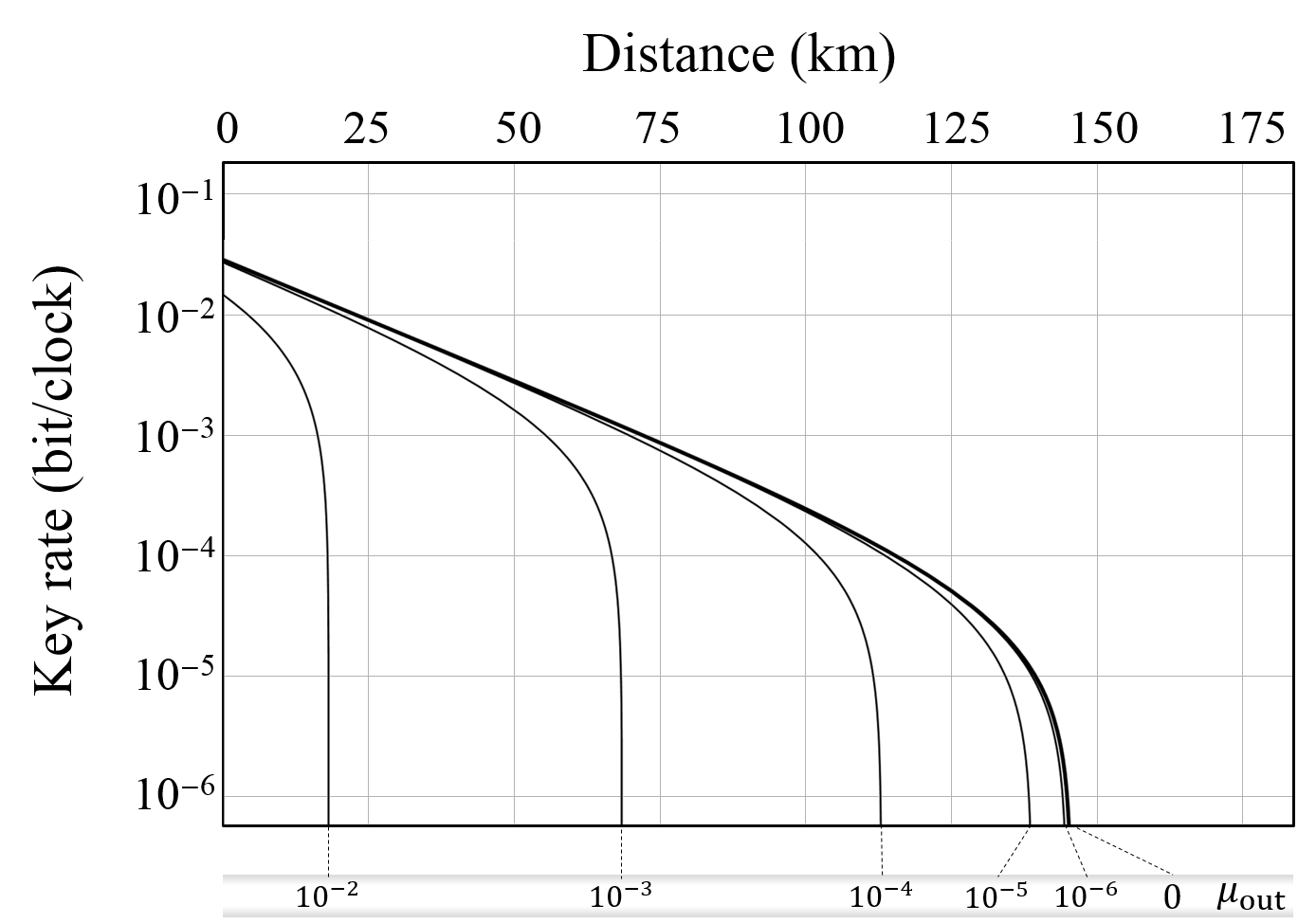}\\
  \caption{Asymptotic key rate $\widetilde{R}^{**}$ versus distance for the decoy-state efficient BB84 protocol, under a THA with unambiguous state discrimination by Eve. The rate is plotted for various values of the parameter $\mu_{\textrm{out}}$. Parameters in the simulation are as in Fig.~\ref{fig:4}.}
  \label{fig:10b}
\end{figure}

In the present THA, whenever the USD provides a conclusive outcome, Eve forwards Alice's pulse to Bob without perturbing it. Therefore only a fraction $1-\delta$ of the counts provide a faithful estimation of the error rate. After bounding the phase error rate as $e'_Y / (1-\delta)$, we can follow similar steps as in Ref.~\cite{GLLP04} to show that secure key bits can be extracted from the single-photon efficient BB84 protocol in presence of the here-described THA at a rate:
\begin{equation}
\label{eq:9bistris}
     R^{**} = Q_X \left\{  \left(  1-\delta \right)  \left[  1-h\left(  \frac{e'_Y}{1-\delta}\right)  \right]  -f_{EC}h\left(  e_{X}\right)  \right\}.
\end{equation}
Eqs.~\eqref{eq:del} and~\eqref{eq:9bistris} can be easily generalized to the case of a decoy-state source under Assumption~3 of Appendix~A:
\bea \label{eq:12bistris}
\nonumber   \widetilde{R}^{**} &=& \widetilde{Q}_X^{(1)} (1-\tilde{\delta}^{(1)})  \left[ 1-h \left( \frac{\tilde{e}^{\prime(1)}_Y}{ 1-\tilde{\delta}^{(1)}} \right) \right] - Q_X^{(s)} f_{EC} h[e_X^{(s)}] ,\\
   \tilde{\delta}^{(1)} &=& \frac{1-\exp(-2\mu_{\textrm{out}})}{\widetilde{\mathcal{Y}}^{(1)}}~,
\eea
with $\widetilde{\mathcal{Y}}^{(1)}=\min[\widetilde{\mathcal{Y}}^{(1)}_X,\widetilde{\mathcal{Y}}^{(1)}_Y]$. In Eq.~\eqref{eq:12bistris}, the tilde indicates quantities to be estimated via the decoy state technique. $Q_X^{(s)}$ and $e_X^{(s)}$ are the same as in Eq.~\eqref{eq:12bis}. The expression of the phase error rate $\tilde{e}^{\prime(1)}_Y$ is the same as in Eq.~\eqref{eq:12bisbis}, because it is estimated in the $Y$ basis which, in this specific THA, does not allow Eve to selectively modify the transmission rate conditional on her information on Alice's state.

The key rates in Eqs.~\eqref{eq:9bistris} and~\eqref{eq:12bistris} are plotted in Figs.~\ref{fig:10a} and ~\ref{fig:10b}, respectively. Despite they are better than the key rates in Figs.~\ref{fig:3} and~\ref{fig:4}, drawn for the most general THA from the GLLP proof method~\cite{GLLP04}, there is no wide gap between the two situations. This suggests that the particular THA described here catches the main features of the general attack described in Sec.~\ref{sec:2}. It also suggests that the real-time use of the auxiliary system $E$ is the main source of troubles in a THA. This seems to be particularly detrimental in the framework of Ref.~\cite{TCK+14}, which heavily relies on Eve's inability to change the transmission rates of the states emitted by Alice.

It would be interesting to extend the proof method of Ref.~\cite{TCK+14} to more general Trojan-horse attacks than the one described in this section. However, such a generalization is not straightforward and a separate analysis is required~\cite{NOTE10}.


\begin{thebibliography}{99}
\bibitem{NOTE1}     Most of the results hold for the receiver side too, but the connection with security depends on the particular protocol used. In the standard BB84 protocol, the receiver is already secure against the THA, see~\cite{VMH01}.
\bibitem{NOTE2}     By definition, LIDT establishes a point where zero damage is caused to the optical component. Hence, the resulting damage threshold is a loose bound. A tighter bound would be one where damage is certainly caused to the component.
\bibitem{NOTE3}     This is the amount of energy transferred to the medium before it starts dissipating heat. The characteristic dissipation time for fused silica is $100~\mu$s, where we used $D=0.75$~mm$^2/$s for the thermal diffusivity in fused silica~\cite{W03} and the mentioned value of the core area.
\bibitem{NOTE4}     The laser used in~\cite{KB88} featured $\lambda=1.064~\mu$m. The corresponding LIDT at $\lambda=1.55~\mu$m is only a factor 1.2 larger, in line with the square root dependence of LIDT on wavelength \cite{CRD03}.
\bibitem{NOTE8}     It would be useful to have an optical fuse at less than 300~mW, which is a common tolerance level specified for fiber-based components. However, despite some recent advertisement, to our knowledge such a device is not yet commonly available.
\bibitem{NOTE5}     Although we are in the asymptotic scenario, where error correction can in principle reach the Shannon limit ($f_{EC}=1$), we follow~\cite{LMC05} in maintaining a factor $f_{EC}>1$  in the rate equation so to provide a more realistic prevision on the final rate.
\end{thebibliography}
\end{document}